\documentclass{aa}  

\usepackage{graphicx}
\usepackage[english]{babel}
\usepackage{amsmath}
\usepackage{natbib}
\usepackage[colorlinks=true, allcolors=blue]{hyperref}
\usepackage{xcolor}
\usepackage{lscape}
\usepackage{pdflscape}
\usepackage{multicol}
\usepackage{txfonts}

\begin{document}

   \title{UV FeII emission model of HE 0413-4031 and its relation to broad-line time delays}
   \titlerunning{UV FeII emission model}

 %  \subtitle{I. Overviewing the $\kappa$-mechanism}

   \author{Michal Zaja\v{c}ek
          \inst{1}\fnmsep\thanks{based on observations made with the Southern African Large Telescope (SALT)}
          \and
          Swayamtrupta Panda\inst{2}\thanks{CNPq Fellow}
          \and
          Ashwani Pandey\inst{3}
          \and 
          Raj Prince\inst{3}
          \and
          Alberto Rodr\'iguez-Ardila \inst{2,4}
          \and
          Vikram Jaiswal\inst{3}
          \and
          Bo\.zena Czerny \inst{3}
          \and
          Krzysztof Hryniewicz \inst{5}
          \and 
          Maciej Urbanowicz \inst{6}
          \and 
          Piotr Trzcionkowski \inst{6}
          \and
          Marzena \'Sniegowska \inst{7}
          \and
          Zuzanna Fałkowska \inst{8,3}
          \and 
          Mary Loli Mart\'inez-Aldama \inst{9}
          \and 
          Norbert Werner \inst{1}
          }

    \institute{Department of Theoretical Physics and Astrophysics, Faculty of Science, Masaryk University, Kotl\'a\v{r}ska 2, 611 37 Brno, Czech Republic
    \and
    Laboratório Nacional de Astrofísica, MCTI, Rua dos Estados Unidos, 154,  37504-364 Itajubá, MG, Brazil
    \and 
    Center for Theoretical Physics, Polish Academy of Sciences, Al. Lotnik\' ow 32/46, 02-668 Warsaw, Poland
    \and
    Instituto Nacional de Pesquisas Espaciais, Divisão de Astrofísica. Avenida dos Astronautas 1758. São José dos Campos, 12227-010, SP, Brazil
    \and
    National Centre for Nuclear Research, ul. Pasteura 7, 02-093 Warsaw, Poland
    \and
    Astronomical Observatory, University of Warsaw, Al. Ujazdowskie 4, 00-478 Warsaw, Poland
    \and
    School of Physics and Astronomy, Tel Aviv University, Tel Aviv 69978, Israel
    \and
     Faculty of Physics, University of Warsaw, ul. Pasteura 5, 02-093 Warsaw, Poland
    \and
    Astronomy Department, Universidad de Concepción, Casilla 160-C, Concepción 4030000, Chile\\
            }         

  \date{Received October 5, 2023; accepted December 20, 2023}

% \abstract{}{}{}{}{} 
% 5 {} token are mandatory
 
  \abstract
  % context heading (optional)
  % {} leave it empty if necessary  
  {FeII emission is a well-known contributor to the UV spectra of active galactic nuclei and the modeling of this part may affect the results obtained for the Mg~II~$\lambda2800$ emission, which is one of the lines used for black hole mass measurements and cosmological applications.}
  {We tested different FeII emission models when modeling the UV emission of the intermediate-redshift quasar HE 0413-4031 to see how the use of a specific template affects the Mg~II~$\lambda2800$ line properties and the measurement of the Mg~II~$\lambda2800$ and UV Fe~II time delays with respect to the continuum.}
  {We used the 11-year monitoring of the selected quasar HE 0413-4031 with the South African Large Telescope (SALT), and we supplemented this monitoring with the near-IR spectrum taken with the SOAR telescope, which gave access to the H$\beta$~$\lambda4861$ and [OIII]~$\lambda \lambda 4959,5007$ emission lines at the rest frame and allowed for a precise measurement of the redshift.}
  {A new redshift determination ($z=1.39117 \pm 0.00017$) using [OIII]~$\lambda \lambda 4959,5007$ gave a very different value than the previous determination based only on the UV FeII pseudocontinuum ($z=1.3764$). It favors a different decomposition of the spectrum into MgII and UV FeII emissions. The line characteristics and the time delay of the MgII emission ($224^{+21}_{-23}$ days) are not significantly affected. However, in comparison with the previous analysis, the rest-frame UV FeII time delay ($251^{+9}_{-7}$ days) is consistent with the inferred UV FeII line full width at half maximum (FWHM) of $4200\,{\rm km/s}$ that is only slightly smaller than the MgII line FWHM. Hence the FeII-emitting material is more distant than the MgII-emitting gas in HE 0413-4031 by $\sim 0.023$ pc (4700 AU). The inferred velocity shift of both MgII and UV FeII lines with respect to the systemic redshift is now rather low, below 300 km s$^{-1}$. In addition, we constructed an updated MgII radius-luminosity ($R-L$) relation from 194 sources, which is more than double the previous sample. The MgII $R-L$ relation is flatter than the UV FeII, optical FeII, and H$\beta$ $R-L$ relations. While the new decomposition of the spectrum is satisfactory, we see a need to create better FeII templates using the newest version of the code \texttt{CLOUDY}.}
  {}

  % conclusions heading (optional), leave it empty if necessary 


   \maketitle
%
%-------------------------------------------------------------------

\section{Introduction}

The characteristic property of the bright active galactic nuclei (AGN) is the occurrence of broad and strong emission lines in the optical and UV bands \citep[see][for reviews]{krolik_book1999,2021bhns.confE...1K,2023Ap&SS.368....8C,2023EPJD...77...56C}. Those emission lines are clearly visible in the individual objects as well as in the quasar composite spectra \citep[e.g.,][]{cristiani1990,francis1991,vandenberk2001}. The strongest emission lines usually listed in the quasar catalogs for individual objects are H$\beta$~$\lambda4861$, MgII~$\lambda 2800$, CIII]~$\lambda 1908$, CIV~$\lambda 1549$, and Ly$\alpha$~$\lambda 1215$, depending on the object redshift and the coverage in the corresponding rest frame. Many more lines are listed in a single source or smaller sample studies \citep[see, e.g.,][]{martinez2018,Horne2021, 2023arXiv230501014S}. Apart from the well-specified lines, there is a contribution from the Balmer continuum supplementing the emission of the accretion disk responsible for the continuum emission. The last but very important contributor is the emission from the FeII and FeIII ions \citep[see e.g.,][for a short review]{gaskell2022}. Numerous transitions present in the optical and UV data lead to the formation of the FeII pseudo-continuum. Particularly, in the vicinity of the MgII line, there is a strong and overlapping contribution from FeII. The issue has been studied for many years \citep[e.g.,][]{collin_souffrin1979,wills1980,WNW1985}. The presence of FeII in bright AGN is universal; a similar FeII/MgII line ratio is observed in AGN at all redshifts, from nearby sources to $z \sim 5 $ objects \citep{dietrich2003}.

However, the FeII contribution to the optical band is not well correlated with FeII in the UV domain \citep[][hereafter also denoted as KDP15]{KDP2015,2019MNRAS.484.3180P}. The optical FeII strongly correlates with the source Eddington ratio, which underlies the concept of the quasar optical main sequence \citep{boroson1992,Sulentic2000,panda2018,panda2019}. The existence of the corresponding UV quasar plane based on MgII line properties and the UV FeII is not so well established \citep{UVplane2020}. The UV and optical FeII emission may be well correlated in a single object but it does not seem to come from the same part of the broad line region (BLR) \citep{Panda_etal_2019_WC, Zhao2020}. In multi-object statistical samples, there is an overall correlation between the quasar luminosity and the UV FeII strength \citep{clowes2016}. Several works \citep{marziani2003,Shen_Ho_2014, panda2019, zheng2021, Panda_Marziani_2023FrASS} claim that the FeII strength can actually be used as the Eddington ratio indicator.

The strong overlap between the UV FeII and MgII emissions complicates the measurements of the MgII time delay and makes the modeling of the emissivity and the dynamics of UV FeII difficult. The FeII models in this spectral range are either constructed from the data, with objects of extremely narrow emission lines, such as I Zw 1, serving as the best templates  \citep[][hereafter also denoted as VW01]{Vestergaard2001}, or theoretically \citep{Bruhweiler2008,KDP2015}. The analysis is additionally complicated by the fact that the FeII emission can be considerably shifted with respect to the rest frame, which can also make an identification of the individual transitions in the data problematic.

The precision of the spectral decomposition close to the MgII line has several important consequences. First, the measurements of the kinematic width of the FeII and MgII components, and the measurements of the corresponding delays should shed light on the stratification of the BLR. The next issue is the proper determination of the radius-luminosity (R-L) relation for the MgII line which is essential for the black hole mass measurements from a single spectrum in large statistical samples \citep[e.g.,][]{wang2009,kdp2017,Le2020}. These R-L relations for both MgII and FeII can also reveal stratification of the BLR. Proper determination of the FeII  is also important from the point of view of the statistical studies of the AGN properties. The principal component analysis (PCA) approach by \citet{boroson1992} clearly showed the dominant role of the two parameters: (i) the ratio of the FeII equivalent width in the optical band to the equivalent width of H$\beta$ line, and (ii) the full width at half maximum (FWHM) of H$\beta$ line. These two parameters form the optical plane, but in a close analogy, a UV plane can be introduced based on FeII emission in UV and MgII line properties \citep{Panda_etal_2019_WC,UVplane2020}. Finally, the R-L relation can be successfully used in cosmological studies to constrain the expansion rate of the Universe \citep{khadka2021}. 

The spectral decomposition into FeII and MgII can be improved by the wavelength-resolved time delay measurements. Such studies of three intermediate-redshift quasars were performed - for CTS C30-10 \citep{2022A&A...667A..42P}, and for HE 0413-4031 and HE 0435-431 \citep{prince2023}, each monitored for 11 years with the Southern African Large Telescope (SALT). The results did not imply any outflow but indicated different apparent stratification of the FeII and MgII emissions in the three quasars. The study, when combined with the separate FeII and MgII time delays available from the literature, also enabled us to create the radius-luminosity relation for both spectral features, which suggests more compact emission regions with FeII-emitting clouds being located closer to MgII-emitting material for a lower luminosity, while at higher monochromatic luminosities, both regions are more distant and they radially converge. However, our results were based on the assumption that FeII is at the systemic redshift and this assumption certainly could bias our choice of the FeII template. 

The redshift determination for the source HE 0413-4031, which is a bright, intermediate-redshift quasar studied in this paper, was recently obtained based on the high-quality single spectrum in the IR band in the observed frame (Panda et al., in preparation). The spectrum covered well the H$\beta$ region with accompanying narrow [OIII] $\lambda\lambda4959,5007$ lines. The position of the [OIII] line allowed for a precise measurement of the source redshift which was considerably different from the redshift used originally by \citet{Zajacek2020} for the MgII line reverberation mapping and by \citet{prince2023} for the wavelength-resolved study. The previous value was determined based on the UV FeII pseudocontinuum transitions seen in the observed optical spectrum. Therefore, in this paper, we study several FeII templates with the final aim of fitting better the spectra and checking whether the substantial change in the FeII/MgII decomposition affects the measurement of the MgII and UV FeII time delays. This could have important implications for the use of the quasar emission-line time delays for cosmology \citep[see e.g.,][for applications]{2019FrASS...6...75P,2021AcPPA.139..389C,khadka2021,zajacek2021,2022MNRAS.516.1721C,2023Ap&SS.368....8C,2023arXiv230508179Z,2023arXiv230916516C}.

The paper is organized as follows. We briefly describe the photometric and spectroscopic data in Section \ref{sec:obs} and present a revised redshift for the source HE 0413-4031. In Section \ref{sect:decompo} we outline the spectral decomposition of our SALT spectra. We present the results (Section \ref{sec:results}) from the spectral decomposition and the time delay analyses from the spectrophotometric data and highlight the significant effect on the measured parameters due to the change in the redshift for this source. We discuss the implication of the new time delays for the MgII and UV FeII emission lines and show the location of HE 0413-4031 in the updated MgII \& FeII R-L relations in Section~\ref{sec:discussions}. The new R-L relation for MgII is based on the largest sample used for this purpose. We also show a new preliminary FeII spectral model using the updated version of the \texttt{CLOUDY} photoionization code, which exhibits several differences with respect to the older model. Furthermore, within the same section, we discuss the location of this source in the UV quasar main sequence. Finally, we summarize our findings in Section  \ref{sec:conclusions}.

\section{Observational data}
\label{sec:obs}

We use almost the same data for the quasar HE 0413-4031, first identified in the Hamburg-ESO survey by \citet{wisotzki2000}, as was done by \citet{prince2023}. We supplement it with one new spectroscopic long-slit measurement from the Southern African Large Telescope (SALT), obtained in January 2023, so the whole spectroscopic monitoring covers the period from January 2013 till January 2023. The basic raw data reduction is done by the standard SALT pipeline for the Robert Stobbie Spectrograph (RSS; \citealt{burgh2003,kobulnicky2003,smith2006}, see also \citealt{2010SPIE.7737E..25C}). 

We supplement the spectroscopic data with photometric data from several telescopes, as described in \citet{prince2023}. This data is needed for the absolute calibration of the spectra.
The basic quasar properties were estimated previously \citep{Zajacek2020}. The black hole mass based on the reverberation mapping and the spectral energy distribution fitting is in the range $\log{(M_{\bullet}\,[M_{\odot}] )}= 9.0-9.7$ \citep{Zajacek2020}. The SMBH accretes close to the Eddington limit, $\lambda_{\rm Edd} = 0.38-2.18$, which was estimated using the inferred range of SMBH masses and different values of the bolometric correction \citep[see][for details]{Zajacek2020}. The quasar is bright, with the $V$ magnitude typically about 16.5 as reported by \citet{veron_cetty2010}. 

\begin{figure}
    \centering
    \includegraphics[width=\columnwidth]{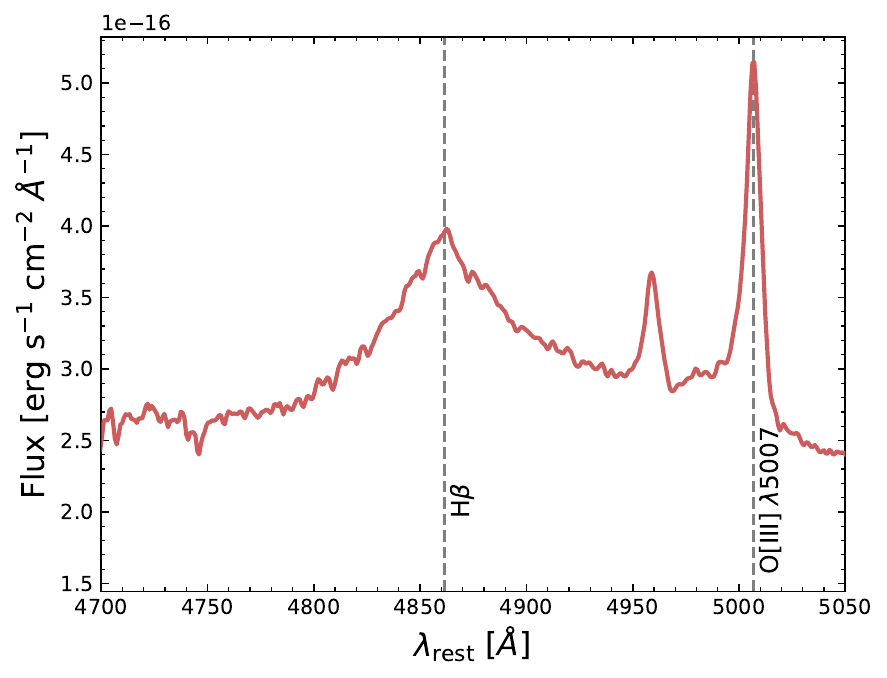}
    \caption{    
    Rest-frame spectrum for the HE 0413-4031 covering the spectral range around the H$\beta$ region obtained using the SOAR/TripleSpec4 NIR spectrograph. The redshift  ($z = 1.39117 \pm 0.00017$) was estimated using the [O {\sc iii}]$\lambda$5007 emission line. The zero-velocity positions of H$\beta$ and [O {\sc iii}]$\lambda$5007 emission lines are marked using the dashed vertical lines.
    }
    \label{fig:optical-spectrum}
\end{figure}

The redshift was estimated as $z=1.3764$ assuming that the UV FeII emission in this source is at the quasar rest frame \citep{Zajacek2020}. Then the best fitting FeII template was \texttt{d12-m20-20-5.dat} (hereafter d12), one of the theoretical templates of \citet{Bruhweiler2008}, corresponding to the cloud number density of $10^{12}\,{\rm cm^{-3}}$, the turbulent velocity of $20\,{\rm km  s^{-1}}$,  and the hydrogen ionizing photon flux of $10^{20.5}{\rm cm^{-2}s^{-1}}$. However, a recent high-quality observation of this source in the near-IR (NIR) using the TripleSpec4 long-slit spectrograph \citep{Schlawin_etal_2014_TSpec4} on the 4.1m SOAR (see Figure~\ref{fig:optical-spectrum} and Table~\ref{tab:SOAR}) brought the new determination of the redshift of $1.39117 \pm 0.00017$ (Panda et al., in preparation). Briefly, we use a custom script that calculates the barycenter of the data points around a selected wavelength range and informs the wavelength with respect to the peak position of the [OIII]$\lambda$5007 emission line. We extract the wavelengths for the doublet [OIII]$\lambda\lambda$4959, 5007 lines separately and find that the resulting redshifts are identical up to the 4th decimal place. Henceforth, we consider the value of the redshift derived from the [OIII]$\lambda$5007 line as quoted above. The error on the redshift is estimated from the RMS of the skylines in the spectrum in the 6th order of the cross-dispersed spectrum, that is where the H$\beta$ region including the [OIII]$\lambda\lambda$4959, 5007 are observed.

The observations were made in cross-dispersed mode and the spectral resolution (R) for this spectrum was $\sim$3500 across the different dispersion orders. Observations were made nodding in two positions along the slit. Right before the science target, a telluric star (HIP18437), close in airmass to the former, was observed to remove telluric features and to perform the flux calibration. Cu-Hg-Ar arc frames were also observed at the same position as the science target for wavelength calibration. The rms for the calibrated spectrum was found to be 0.201 \AA\ and the wavelength scale has been realigned to the wavelength of the skylines as recorded on the spectrum of the target. The spectral reduction, extraction, and wavelength calibration procedures were performed using {\sc spextool} v4.1, an IDL-based software developed and provided by the SpeX team \citep{Cushing_etal_2004PASP} with some modifications specifically designed for the data format and characteristics of TripleSpec4, written by Katelyn Allers (private communication). Telluric feature removal and flux calibration were done using {\sc xtellcor} \citep{Vacca_etal_2003PASP}. The different orders were merged into a single 1D spectrum from 1 to 2.4 microns using the {\sc xmergeorders} routine.

This new redshift value implies a rough velocity shift by 4400 km s$^{-1}$. It implies that either the FeII is outflowing with an extreme velocity, higher than typically seen in the UV FeII component (below 3000 km s$^{-1}$; \citealt{KDP2015}), or the FeII template used in our previous work is not correct. For this reason, we try out different UV FeII templates to see if we can obtain a better fit and overall consistency than previously.
%by \citet{prince2023} is not correct. 

%-----------------------------------------------------------------
   \begin{table}
      \caption[]{SOAR IR fits with redshift $z = 1.391 17 \pm 0.000 17$ implied by fitting narrow lines. The full width at half maximum (FWHM), the line dispersion ($\sigma$), and the line equivalent width (EW) are reported for the broad and the narrow components of the H$\beta$, and the narrow components for the [O {\sc iii}]$\lambda$4959 and [O {\sc iii}]$\lambda$5007, respectively.}
         \label{tab:SOAR}
         \begin{tabular}{lccc}
         \hline
         \hline
            \noalign{\smallskip}
            Component      &  FWHM & $\sigma$ & EW\\
                        &  [km s$^{-1}$] & [km s$^{-1}$] & [\AA] \\
            \noalign{\smallskip}
            \hline
            \noalign{\smallskip}
    Broad H$\beta$ &  5457$\pm$238 &  2820$\pm$21 &  55.1$\pm$0.7 \\
    Narrow H$\beta$ &  552.3$\pm$24.1 &  229.2$\pm$1.7 &  0.70$\pm$0.01  \\
    Narrow [OIII]4959 &  552.3$\pm$24.1 &  229.2$\pm$1.7 &  3.60$\pm$0.04  \\
    Narrow [OIII]5007 &  552.3$\pm$24.1 &  229.2$\pm$1.7 &  9.20$\pm$0.11  \\
            \noalign{\smallskip}
            \hline
         \end{tabular}
   \end{table}
%

%----------------------------------------------------------------- 

\section{Spectral decomposition method}
\label{sect:decompo}

The quasar HE 0413-4031 accretes at the high Eddington ratio and it shares the characteristics of type A quasars \citep{Sulentic2000}, analogs of Narrow Line Seyfert 1 class \citep{osterbrock1985} for lower-mass sources, especially its strong emission lines are well fitted with a single Lorentzian profile. Although the FWHM of the broad H$\beta$ line is $5457 \pm 238\,{\rm km\,s^{-1}}$ (see Table~\ref{tab:SOAR}), which is above the canonical FWHM division of $4000\,{\rm km\,s^{-1}}$ between type A and B quasars \citep{Sulentic2000}, this division shifts toward a larger FWHM for high-luminosity quasars as HE 0413-4031 \citep[see][for discussion]{Marziani_etal_2018FrASS}.  

Thus, following the previous studies on this source, we assume a single kinematic component for the MgII line shape \citep{Zajacek2020,prince2023}. However, we take into account that MgII is a doublet, and this is included in our modeling, assuming the same widths for the two components, fixed separation. The value of the doublet ratio was optimized by \citet{Zajacek2020} using the mean spectrum. We use the same value of 1.9 in the current paper. The kinematic width, line position, and normalization are free parameters of the model. The underlying continuum is described as a power law with an arbitrary normalization and a slope. We fit the data in a relatively narrow range to avoid issues with broad-band normalization of the SALT spectra. In \citet{prince2023}, we used the rest-frame range of 2700 - 2900 \AA~ which corresponds to 6416.50 - 6891.79 \AA~ in the observed frame. Next, we aim to compare fits for the two values of the source redshift. Here we fit the data in the same observed wavelength range, corresponding to fitting the data in the range of 2683.09  -  2881.84 \AA~ in the rest frame for the new redshift of $1.39117\pm 0.00017$. This is important since there is imperfect subtraction of the sky lines at the longest wavelength tail which might affect the data fitting and proper comparison of the results.

For the FeII pseudocontinuum, we test several templates: theoretical templates of \citet{Bruhweiler2008}, observational VW01 templates, and mixed KDP15 templates. Templates of \citet{Bruhweiler2008} are just a collection of atomic transitions, so they require two parameters: template broadening and overall normalization. The VW01 template requires also two parameters for additional broadening. KDP15 templates contain six components with independent normalizations (five atomic representative transition groups - Multiplets 60, 61, 62, 63 and 78, plus I Zw 1 component), so overall they require 7 parameters. Optionally, we also consider the presence of the He II line at $2733.28\AA$ in the quasar spectrum. 

\section{Results}
\label{sec:results}

In this section, we assess the best FeII model template for the precisely determined redshift of HE 0413-4031, and subsequently, we analyze the impact of the selected FeII templates on the MgII and FeII emission full-width at half maximums (FWHMs), equivalent widths (EWs), light-curve properties, and time delays with respect to the continuum emission. 

\subsection{Spectral decomposition and the best FeII template}

\begin{figure}
   \centering
  \includegraphics[width=0.95\linewidth]{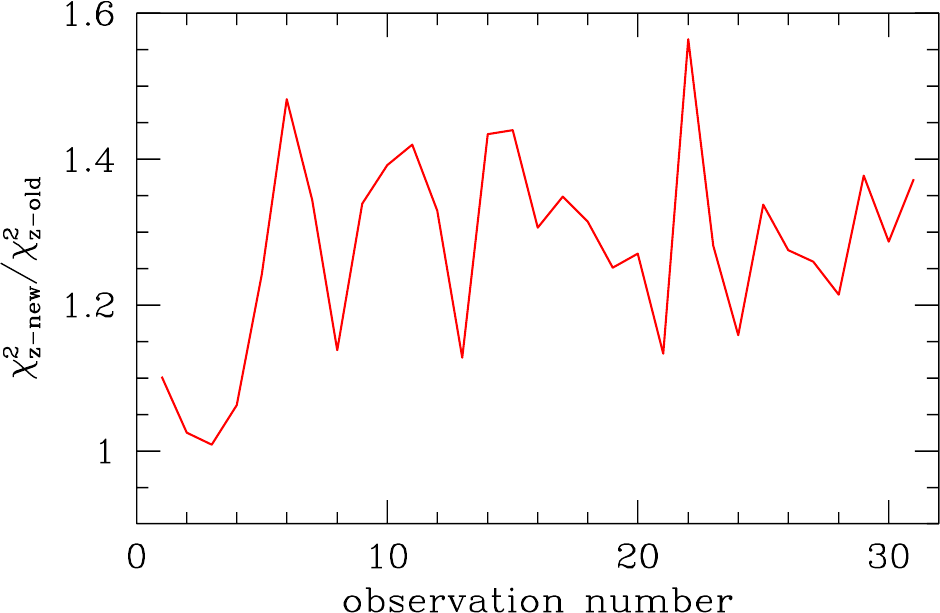}
      \caption{Ratio of the new to the old value of $\chi^2$ when two redshifts for the FeII template d12 are used.
              }
         \label{fig:ratio_chi2}
   \end{figure}

In the previous papers \citep{prince2023, Zajacek2020}, we used the d12 template of \citet{Bruhweiler2008}. Thus, after the redshift correction, we first fit the data with the same template. However, all the fits with the d12 template for the new redshift are worse. The ratio of the new $\chi^2$ value to the old one is shown in Figure~\ref{fig:ratio_chi2}. The previously used FeII template is thus not compatible with the new redshift measurement.

%----------------------------------------------------------------- 
%-----------------------------------------------------------------

\begin{figure}
\centering
\includegraphics[scale=0.4]{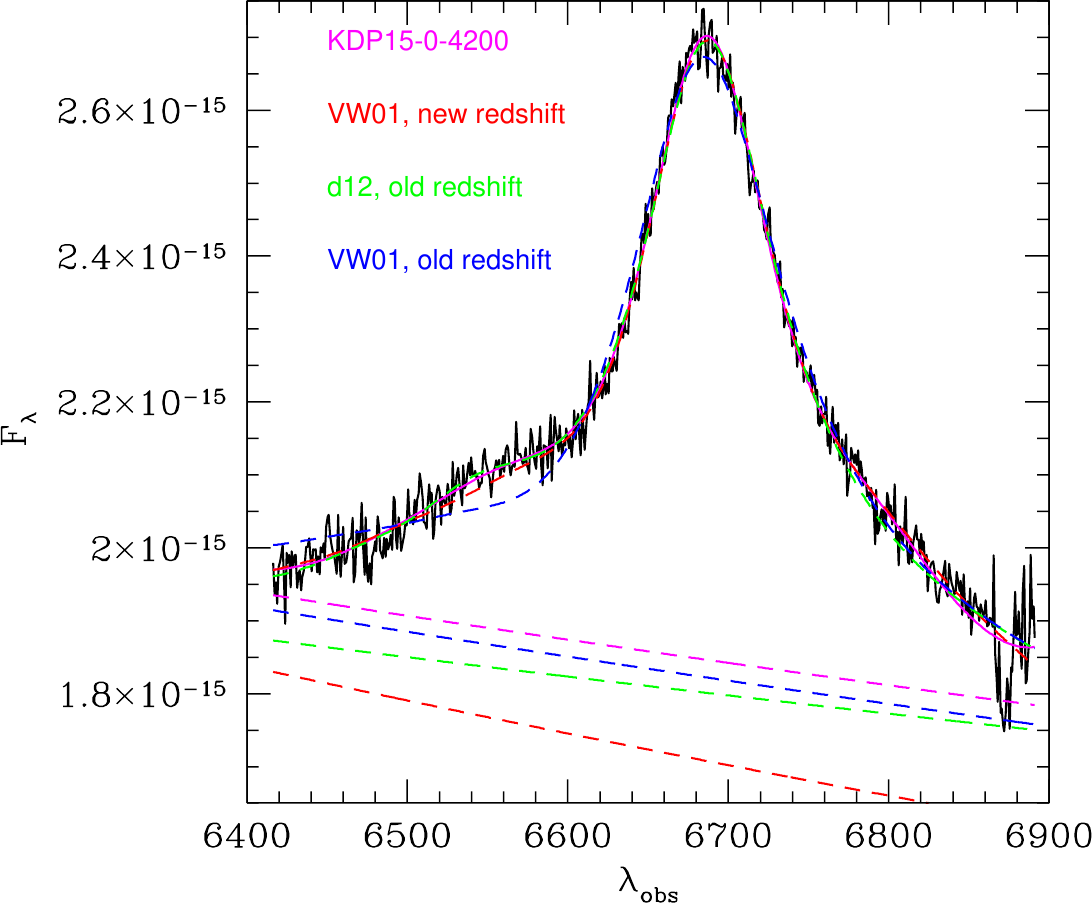}
\includegraphics[scale=0.4]{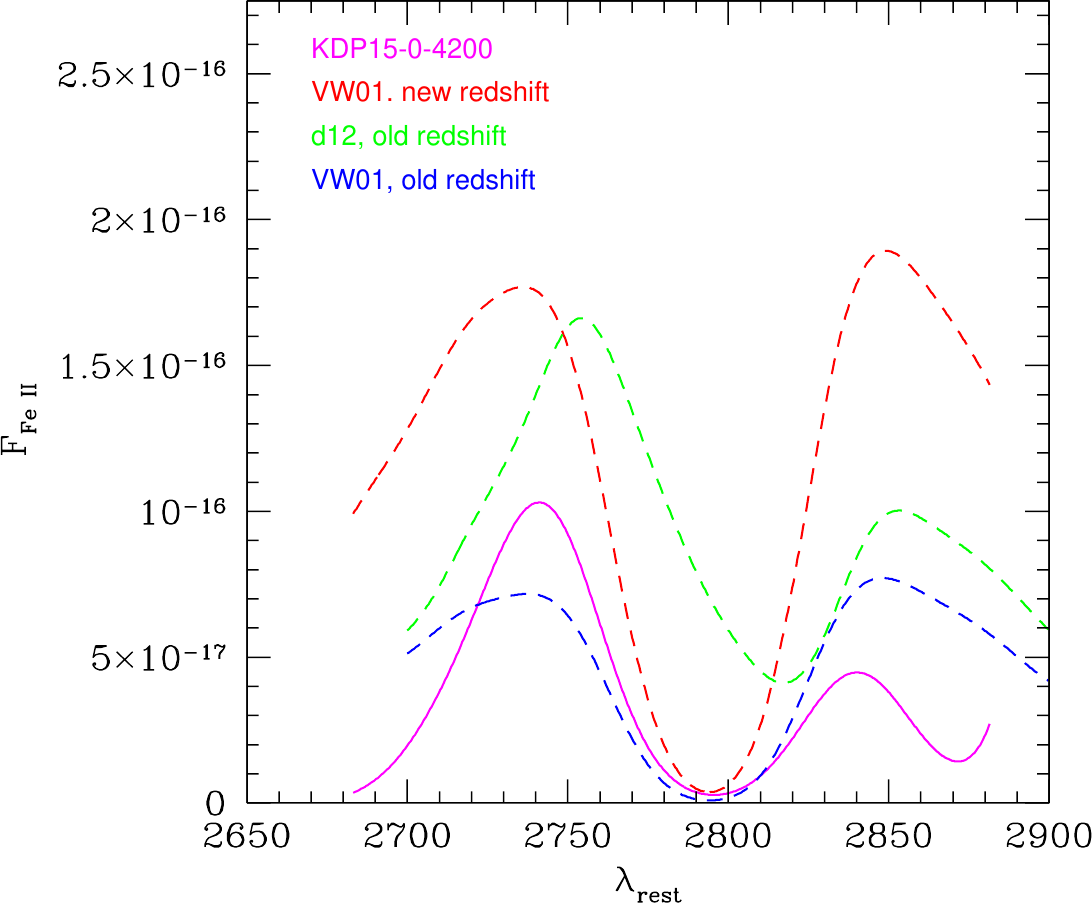}
\includegraphics[scale=0.4]{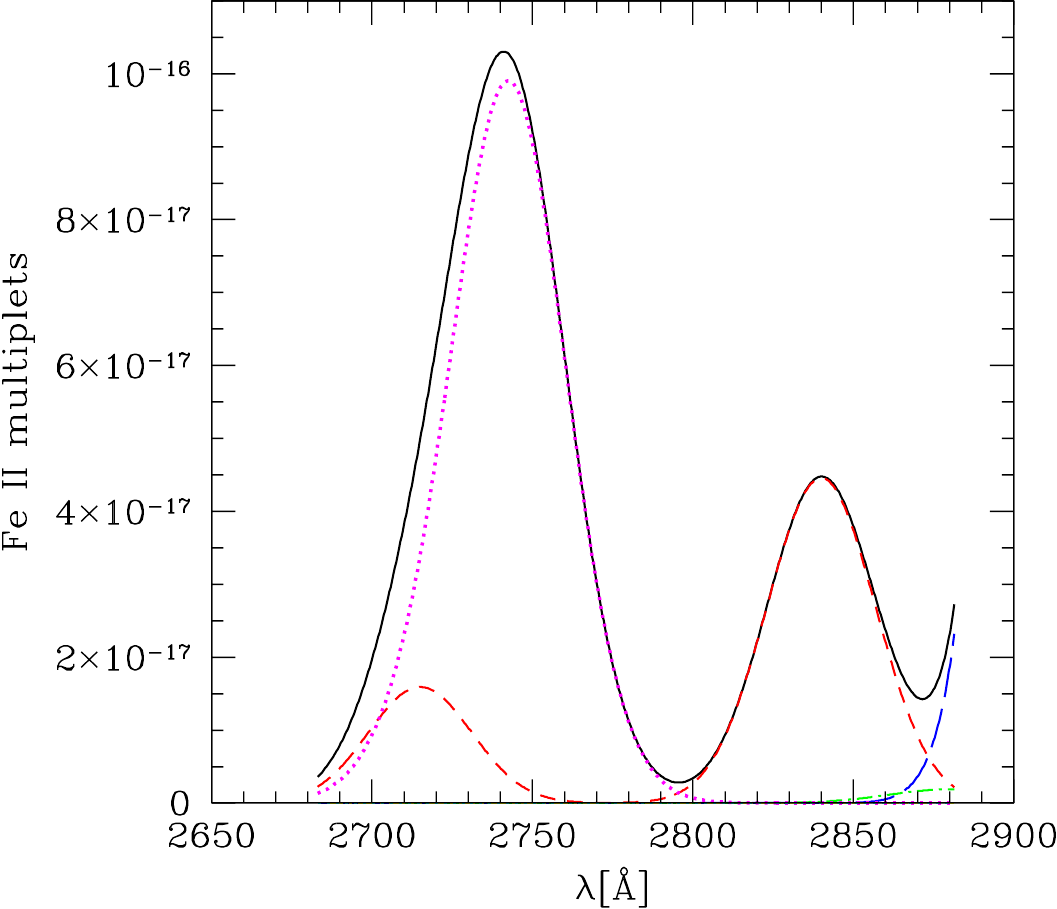}
\caption{Modeling spectra with various FeII templates for the old and the new redshift. Upper panel:  Four fits of Observation 14 in the observed frame: two models for the old redshift $z = 1.37648$, d12 and VW01 templates with extra broadening, and two for the new redshift $z = 1.39117$, the KDP15 template, shift = 0, FWHM = 4200 km s$^{-1}$, and the VW01 template with additional broadening. Dashed lines show the corresponding continua. Middle panel: FeII contribution in the rest frame for all the FeII templates used for the fits in the upper panel (see the legend). Lower panel: Decomposition of the fitted FeII with the KDP15 template into individual multiplets: I Zw 1 (red dashed line), multiplet 60 (blue long dashed), multiplet 61 (green, dash-dot), multiplet 62 (yellow, invisible in the scale), multiplet 63 (magenta, dotted), and multiplet 78 (cyan, almost invisible in the scale), with the black line corresponding to the total FeII emission. The FeII flux density in all three panels is expressed in erg s$^{-1}$ cm$^{-2}$ \AA$^{-1}$.}
\label{fig:new_redshift}
\end{figure}

The issue is a suitable representation of the apparent excess which is now seen at $\sim$ 2730 \AA, see the spectrum in Figure~\ref{fig:new_redshift} close to 6540\AA\ in the observed frame. With the previous value of the redshift, it was located at $\sim 2750$ \AA~ and it was fitted by one of the strong FeII transitions seen in all \citet{Bruhweiler2008} templates. Thus, we supplemented the d12 FeII template with the possibility of the contribution from the He II line $\lambda2733.28$. This line was observed, for example, in the quasar IRAS P09104+4109 by \citet{tran2000}. We assumed the arbitrary position, the width, and the shift of this additional line. However, this did not improve the fit considerably. We also tested the VW01 template but the fit quality with the new redshift was very low. This could be due to the sharp FeII peaks in this template based on the observed spectrum of I Zw 1 with exceptionally narrow broad emission lines. Thus, we applied additional broadening to this template by convolving it with a Gaussian profile. For a grid of models with the Gaussian width sampled from 1000 to 1400 km s$^{-1}$ with the step 100 km s$^{-1}$, we obtained a $\chi^2$ minimum at 1300 km s$^{-1}$. We give the value of 2600 km s$^{-1}$ in Table~\ref{tab:Fe} since this better represents the FWHM. The new fit not only improved in comparison with the previous case but it became much better than the fit of the old template with an extra He II line.

Next, we used the set of UV FeII KDP15 templates. In the previous paper, we did not favor this template \citep[see][their Table C1]{Zajacek2020} as it required a two-component shape of the MgII line to give an improved fit with respect to the d12 template. However, we were using smaller values of the redshift there. Now, with the new redshift from the IR data, we can indeed fit the data significantly better with this template.

%-----------------------------------------------------------------
   \begin{table*}
      \caption[]{HE 0413-4031 fits of Observation 14  for different decompositions. When the redshift is not mentioned, z = 1.39117 is adopted. Negative velocity implies blueshift and $k$ is the number of free parameters in the model. }
         \label{tab:Fe}
         \begin{tabular}{lcccccccc}
         \hline
         \hline
            \noalign{\smallskip}
            Model      &  FWHM(MgII) & EW(MgII) & MgII shift &  FWHM(FeII) & EW(FeII) & $\chi^2$ & $k$ & $\Delta$ BIC\\
                        &  [km s$^{-1}$] & [\AA] & [km s$^{-1}$] & & [\AA] \\
            \noalign{\smallskip}
            \hline
           \noalign{\smallskip}
z = 1.37648, d12 & 4601.6$^{+105.8}_{-104.6}$ & 28.1$^{+0.3}_{-0.3}$ & 1668$^{+5}_{-24}$ & 2830 & 10.3$^{+0.6}_{-0.6}$ & 1246.04 & 6 & 0\\ % 1284.21 BIC reference
z = 1.37648, VW01 & 5187.5$^{+129.4}_{-114.7}$& 29.8$^{+0.5}_{-0.5}$ & 1472$^{+37}_{-37}$ & - & 4.9$^{+0.6}_{-0.6}$ & 1940.00 & 6 & 693.96\\ 
z = 1.37648, VW01 & 5218.7$^{+110.9}_{-121.6}$ & 30.0$^{+0.6}_{-0.6}$ & 1476$^{+49}_{-34}$ & 2400$*$ & 5.2$^{+0.7}_{-0.7}$ & 1932.50 & 6 & 686.46\\ 
d12              & 4906.3$^{+50.3}_{-121.7}$ & 28.0$^{+0.4}_{-0.4}$ & -67$^{+19}_{-24}$ & 2830 & 5.7 $^{+0.5}_{-0.6}$ &  1787.35 & 6 &541.31 \\
d12 + He II$\lambda$2733.28 & 5046.9$^{+150.0}_{-119.7}$ & 28.9$^{+0.3}_{-0.5}$ & -108$^{+20}_{-34}$ & 2830 & 3.8$^{+0.7}_{-0.7}$ & 1485.76 & 9 & 258.80 \\
VW01 & 5085.9$^{+89.0}_{-115.1}$& 33.7$^{+0.5}_{-0.6}$ & -212$^{+12}_{-22}$ & - & 7.8$^{+0.5}_{-0.5}$ & 1478.00 & 6 & 231.96\\ 
VW01 & 5031.3$^{+115.5}_{-109.2}$& 37.2$^{+1.0}_{-1.2}$ & -237$^{+34}_{-9}$ & 2600$*$ & 13.8$^{+4.0}_{-1.9}$ & 1189.35 & 6 & - 56.69 \\  % Vestergaard
KDP15-0-4200 & 4503.1$^{+58.0}_{-58.9}$ & 27.1$^{+0.2}_{-0.3}$ & -270 $^{+15}_{-15}$ & 4200 & 4.0$^{+0.43}_{-0.43}$ &  1130.50 & 11 &  - 83.74 \\ % bubu9
%AGN-template-5600 & & & & & 2111.57 ????\\ % bubu8
            \noalign{\smallskip}
            \hline
         \end{tabular}
         \\$^*$ additional broadening on top of that originally in the template
   \end{table*}
            
%

%----------------------------------------------------------------- 
   \begin{figure}
   \centering
   \includegraphics[width=\linewidth]{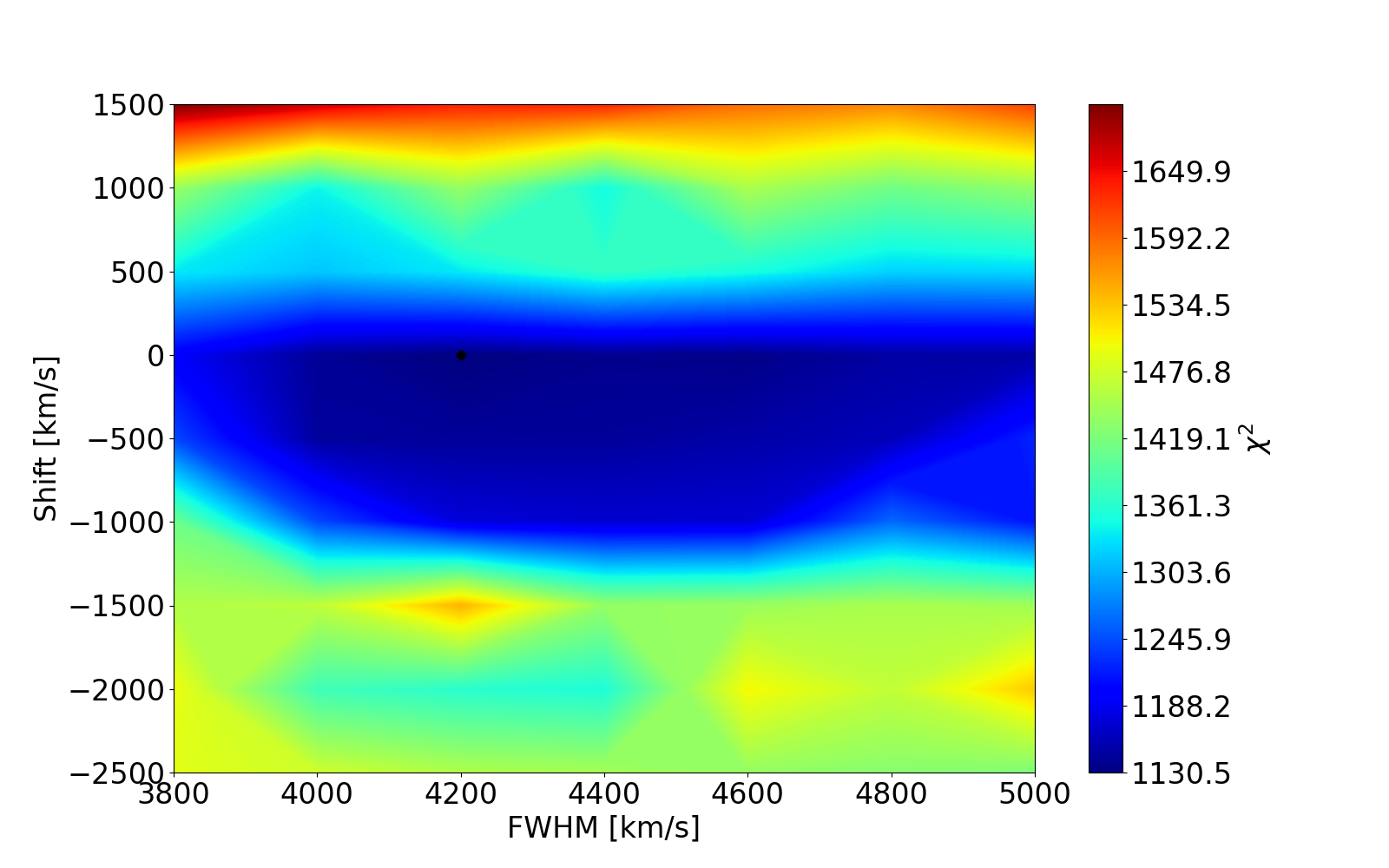}
   \includegraphics[width=\linewidth]{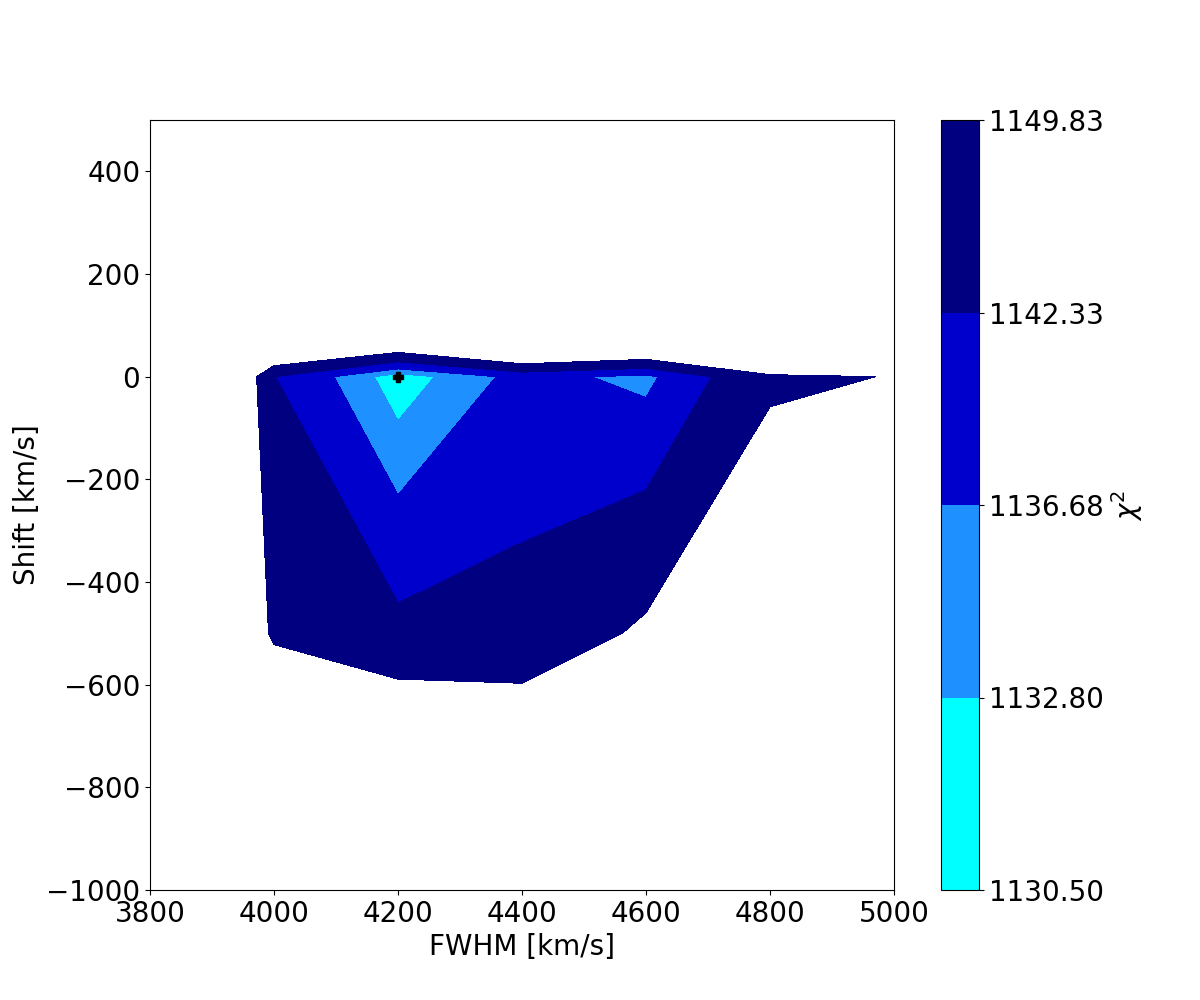}
      \caption{Contour plot for fitting the KDP15 template to Observation 14. The upper panel shows the results for a broad parameter range, and the lower panel shows the best fit, and 1$\sigma$, 2$\sigma$, 3$\sigma$, and 4$\sigma$ errors.
              }
         \label{fig:contour_chi2}
   \end{figure}
%-----------------------------------------------------------------

We compare various spectral fits in detail for Observation 14, which is one of the best from the SALT observing campaign for the studied object (see Table~\ref{tab:Fe}). For the KDP15 template, we studied the width range from 3800 km s$^{-1}$ to 5000 km s$^{-1}$, and the shift range from +1000 km s$^{-1}$ to -4000 km s$^{-1}$. The contour plot is shown in Figure~\ref{fig:contour_chi2}. 
The best fit was achieved for a zero velocity shift of the FeII contribution and the broadening factor of FWHM$ = 4200$ km s$^{-1}$. 
The improvement in the $\chi^2$ value is by more than 100, which is significant for the number of new additional parameters of 5 (we now have 6 independent normalization factors of the FeII multiplets instead of 1 free parameter).  
However, since the number of degrees of freedom in the fits with d12 template and KDP15 template are different, we also determine the Bayesian Information Criterion ($BIC$) for these fits, which in the case of $\chi^2$ fitting method reduces to $BIC = k \log{n} + \chi^2$, where $k$ is the number of free parameters in the model and $n$ is the number of data points (in our case, we have 578 points). The strong decrease in the BIC parameter supports the claim that the fit with the KDP15 template and the new redshift is better.
The fitted parameters are given in Table~\ref{tab:Fe}. 

Visually, both fits, d12 and KDP15 with the shift of $0\,{\rm km\,s^{-1}}$ and the broadening of $4200\,{\rm km\,s^{-1}}$ (denoted as KDP15-0-4200), provide a good fit to the data (see Figure~\ref{fig:new_redshift}, upper panel). However, the decomposition of the spectrum into the MgII component and the FeII pseudo-continuum in both cases is rather different. We plot the FeII component alone in Figure~\ref{fig:new_redshift} (middle panel) with the same normalization. The FeII contribution to the spectrum is much lower for the KDP15 template and its shape in the rest frame is very different. We have made a test of fitting another observation (Observation 7) using again a broad range of shifts and widths for the KDP15 template. The best fit was found for slightly different parameters: the best shift of $-500$ km s$^{-1}$, and FWHM $= 4800$ km s$^{-1}$, although $\chi^2$  at this minimum is not much lower than the $\chi^2$ value for the best shift and FWHM favored for Observation 14.

The properties of the fitted FeII template are also rather surprising since the normalization of the multiplet 62 \citep[see Appendix C in][for the descriptions of individual UV FeII transitions]{Zajacek2020}, usually dominating at 2750 \AA, converged to zero in the fitting process. Instead, most of the FeII flux at the blue wing of MgII (here and in the further text, the ``blue wing'' of the UV FeII refers to the wavelength range 2683.09-2800.00\,\AA\, in the source rest frame while the ``red wing'' refers to the portion at 2800.00-2881.50\,\AA) is now provided by the Multiplet 63 as well as by the I Zw 1 component contribution (see Figure~\ref{fig:new_redshift}, lower panel). It is also surprising to note that the width of the FeII lines in the new fit is only slightly narrower than the FWHM of the MgII line in this observation, which is 4503 km s$^{-1}$. However, this small FWHM difference is consistent with the inferred MgII and FeII time delays, see Subsection~\ref{subsec_time_delay}. 

To test the role of the FeII template shape, we fit Observation 14 using the VW01 template. It gave a good fit when an optimized extra broadening was applied but not as good as the KDP15 template. It resulted in a much stronger FeII contribution, comparable to what we obtained for the old redshift and the d12 template. This template also favored the new redshift, as is seen from Table~\ref{tab:Fe}.

Since the KDP15-0-4200 template was the most satisfactory for the new redshift, we used it to fit all observations. We allowed all six normalization parameters of the template to vary but the width and the shift were the same. We do not expect drastic variations in the properties of the FeII emission during the campaign.

\begin{figure}
   \centering
\includegraphics[width=0.98\linewidth]{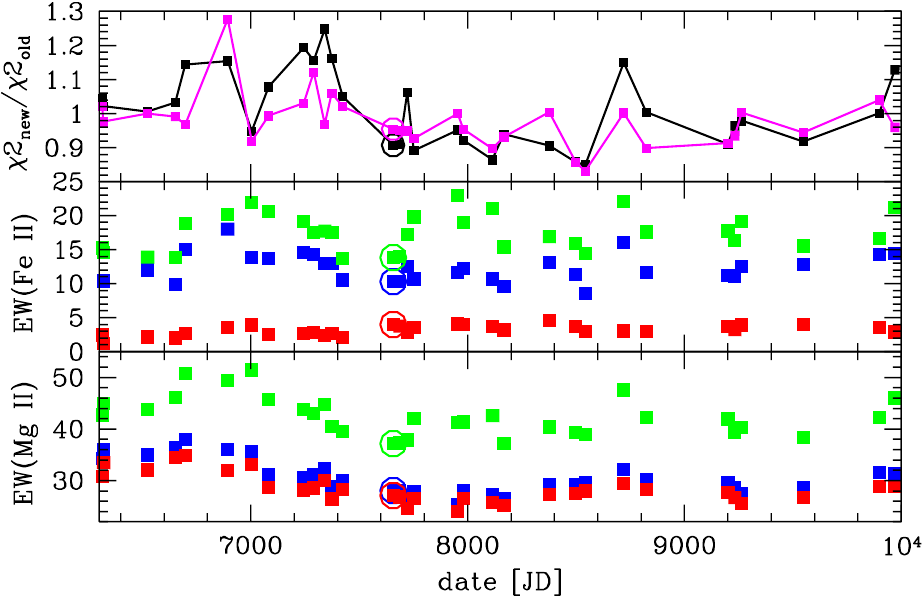} 
      \caption{Comparison of the fit quality using the new UV FeII templates and the previous one. \textit{Upper panel:} $\chi^2$ ratio between the new redshift fits for the KDP15 template with no shift, 4200 km/s broadening (black line), and for the VW01 template (magenta line) with additional broadening and the fit with the d12 template and the old redshift. \textit{Middle panel:} EW(FeII). \textit{Lower panel:} EW(MgII) in these fits (blue, old redshift and template; red, KDP15; and green, VW01). Observation 14 is highlighted with an open circle.
              }
         \label{fig:new_trends1}
   \end{figure}

In Figure~\ref{fig:new_trends1}, we compare the fit quality for all the observations for the new redshift and the new templates (KDP15 and VW01) to the previously used redshift and the old d12 template. We see that some observations are now fitted better and some are not. Statistically, the new redshift and the new template combination works slightly better since the average ratio of the corresponding $\chi^2$ is 0.99. Specifically, the newly selected template with the new redshift gave the average value of the $\chi^2$ of 1793.77 while the fit to all data with the old redshift and d12 template gave 1797.73. Therefore, on average, we do not see a high improvement, particularly when we keep in mind that the fit has, in principle, five more free parameters related to the FeII template. However, the fit is equally satisfactory as before and it uses the new redshift coming from the IR data. 

The new spectral decomposition does not strongly affect the MgII line. The average value of the EW(MgII) is now $28.5 \pm 0.5$\AA\, with a dispersion of 2.8 \AA, while when we use the old redshift and the old template, it is $30.7 \pm 0.6$\AA\, with the dispersion of 3.2 \AA. The small systematic shift is visible in Figure~\ref{fig:new_trends1}, lower panel and the level of variability dropped slightly. The pattern remains roughly the same as in our old analysis, with the initial strong rise, followed by a decrease up to JD 5000 and smaller variations afterward. Only the roughly systematic shift down by $\sim 2$ \AA~is visible. 

The new fit to MgII also shows an interesting improvement after the change of the redshift. The position of the line during the fitting is a free parameter and the average shift of MgII now is only -234 km s$^{-1}$ (blueshift), so the line is much closer to the systemic redshift than in the previous fitting (1650 km s$^{-1}$, redshift). The average width of the MgII line remains larger than that of the FeII (4605 km s$^{-1}$ vs. 4200 km s$^{-1}$). A small systematic shift of the MgII line (from -212 to -270${\rm km\,s^{-1}}$) is consistent with the velocity shifts with respect to [OIII] $\lambda5007$ in the composite spectra of the quasar population \citep{vandenberk2001}. 

However, the properties of FeII drastically changed. The new decomposition lowered the FeII EW level from the average $12.5 \pm 0.4$\AA ~ down to $3.2 \pm 0.1$\AA~ in the studied wavelength band. Thus, the fractional variability in EW(FeII) is now higher, 24 \% while in the previous analysis, it was 17 \%. We did not subtract the measurement error here, so the quoted value is not the excess variance. In addition, the change does not look like a simple shift pattern but shows considerable randomness in time. 

The fitted FeII template in principle has six independent components, corresponding to five Multiplets plus one component named I Zw 1, added for completeness \citep{KDP2015}. This extra component did not find support in the theoretical analysis of the strongest FeII transitions close to MgII but some strong features contributing at $\sim 2715 $ \AA~ and peaking at around $\sim 2840$ \AA~ were present in the spectrum of the object I Zw 1, and such a component seemed to be required to fit the spectra of other objects so it was included in the KDP15 set as a sixth component named \textit{UVFeII-IZw1 lines}. We later refer to this component as I Zw 1 component. However, not all of the six components were actually important for fitting the object studied in this paper. Actually, during the fitting, some of these six components were best fitted with a zero normalization, as in the case of Observation 14. The key element that never vanishes is the Multiplet 63. The second component that is important in some of the spectra - but not in all the spectra - is the empirical I Zw 1 component. The largest variation was seen in Multiplet 60 which was occasionally zero, but sometimes present with a relatively large amplitude. This is related to the fact that this transition is usually strong but it peaks around 2950 \AA, and below 2900 \AA~ we see only its tail \citep[see][their Figure 4]{KDP2015}. A larger number of the degrees of freedom in the FeII template increased the relative variations of the FeII intensity as a function of time. It happened despite the fact of using the same shift and width in all 32 data sets. We additionally tested the effect of the change of the shift and the FWHM of the FeII on the final FeII strength. As we mentioned above, for Observation 7, the best shift was -500 ${\rm km s^{-1}}$, and the best FWHM was 4800 ${\rm km\,s^{-1}}$, which slightly differs from the values adopted here and based on Observation 14. In this nonstandard fit, the EW of FeII was indeed higher, 6.9 \AA~ instead of 3.9 \AA~ for a standard fit illustrated in Figure~\ref{fig:new_trends1}. However, this value is still below the value obtained with the old redshift (13.8 \AA).

\begin{figure}
   \centering
   \includegraphics[width=0.95\linewidth]{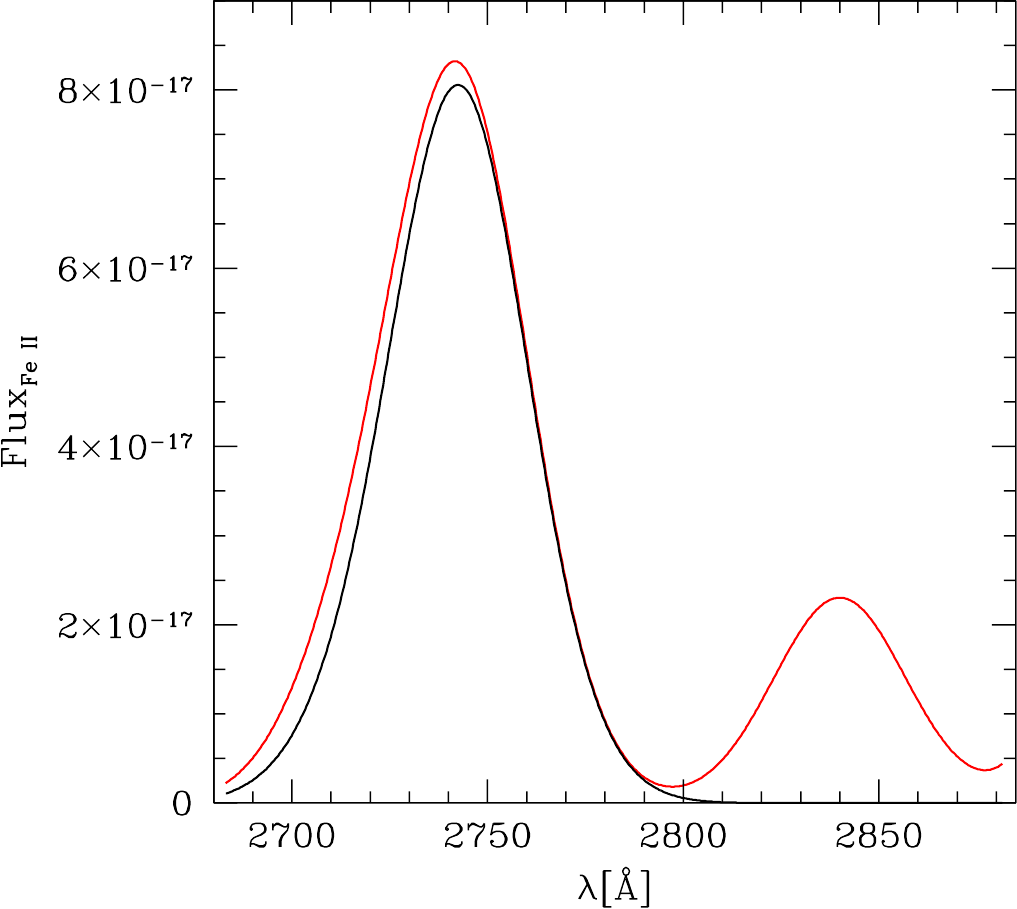}
      \caption{Mean FeII emission averaged over all 32 observations (red) as fitted by all six components. The black line represents the contribution from the multiplet 63 and the remaining part is almost entirely due to the I Zw 1 component (see KDP15), for FWHM of 4200 km s$^{-1}$, and no shift with respect to the systemic redshift. FeII flux density is in units of erg s$^{-1}$ cm$^{-2}$\AA $^{-1}$
              }
         \label{fig:Fe_aver}
   \end{figure}

To illustrate better the FeII properties in the current fits, we created the average FeII profile by combining FeII contributions from all the observations. In Figure~\ref{fig:Fe_aver}, we show the mean unnormalized FeII as fitted by the model, together with Multiplet 63 (black line). This component peaks at $\sim 2742 $ \AA. The weakness of Multiplet 62 is rather surprising since this should be the strongest of the transitions.

%----------------------------------------------------------------- 
%-----------------------------------------------------------------

%-----------------------------------------------------------------

\subsection{MgII and FeII light curves} 

%----------------------------------------------------------------- 
   \begin{figure}
   \centering
   \includegraphics[width=0.95\linewidth]{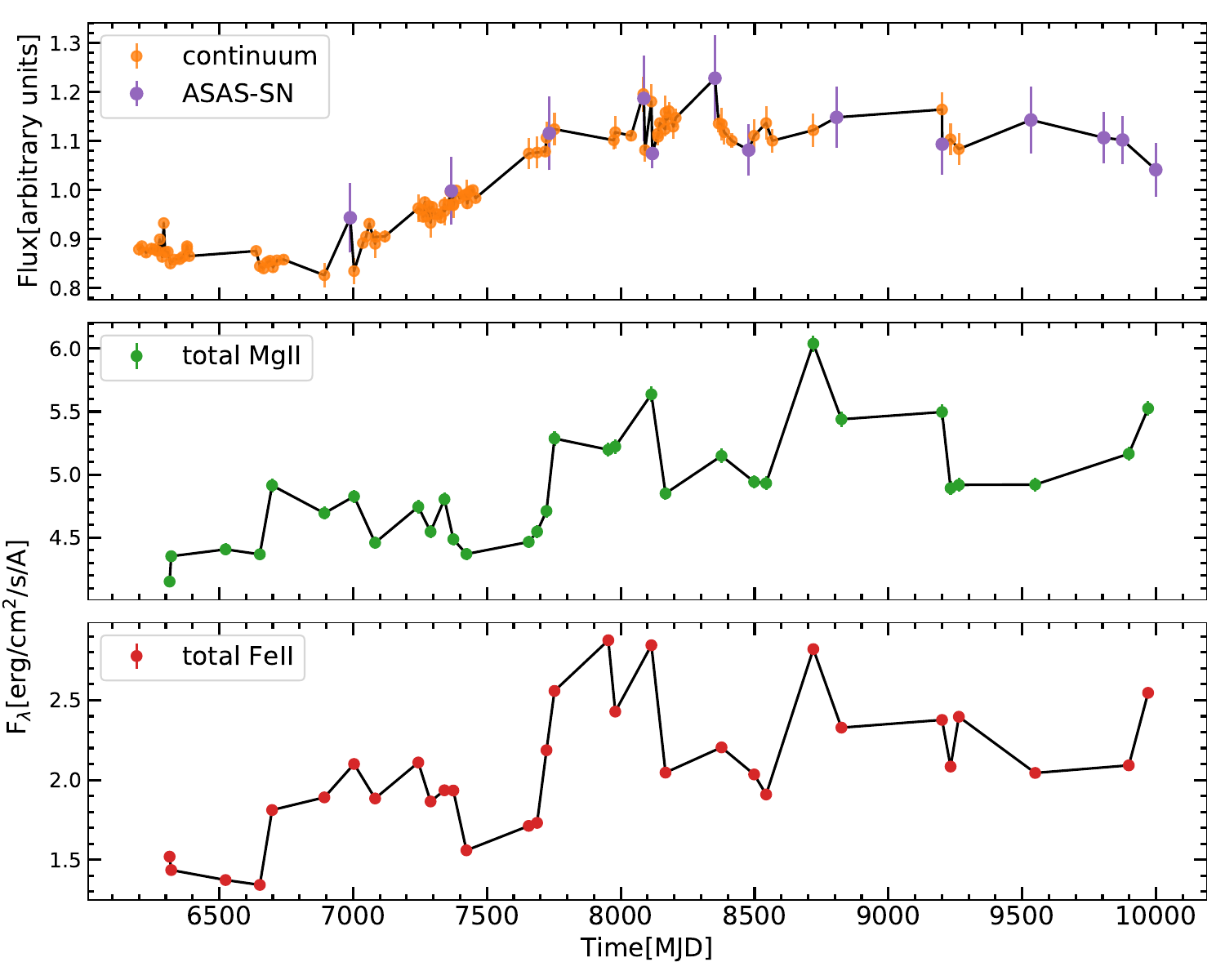}
      \caption{All photometric light curves including archival data from the ASAS-SN. The middle and lower panels show the total MgII and FeII emission light curves in units of 10$^{-16}$ erg cm$^{-2}$ s$^{-1}$ derived using the new redshift, z = 1.39117, respectively. The FeII pseudocontinuum was modeled using the VW01 template.
              }
         \label{fig:curves}
   \end{figure}
%-----------------------------------------------------------------

To perform time-delay measurements, we created MgII and FeII light curves as we did in \citet{prince2023} by normalizing each spectrum with the help of photometry and subtracting the underlying power-law component as well as MgII and FeII contributions, accordingly. In comparison to \citet{prince2023}, in this work, we have an additional SALT spectrum taken on January 26, 2023 (MJD 59970) extending the total duration from January 2013 to January 2023. The SALT spectrum is reduced following a similar procedure as described in \citet{prince2023}. The total length of the continuum light curve is 3799.30 days with a mean cadence of 37.25 days. The BLR line-emission light curves have a total length of 3655.98 days, with a mean sampling of 117.93 days. 

The normalized MgII and FeII light curves, together with the continuum light curve are shown in Figure~\ref{fig:curves}. For the continuum $V$ band data, the photometric information is collected from various telescopes, namely the 1.5 m telescope at OGLE observatory, 40 cm
Bochum Monitoring Telescope (BMT), and SALT measurement with the SALTICAM (all are shown in orange color). In addition, we also include the archival $V$-band data from ASAS-SN (in purple) to extend the baseline of the continuum emission.

%/home/bcz/work/papers/HE_0413_2019/best_fits old results
% /home/bcz/work/papers/HE_0413_2019/best_fits_new_redshift_new_templates/all_observations_Serbia_4200_0 new results

\subsection{MgII \& FeII time delays}
\label{subsec_time_delay}

Using the continuum, MgII, and FeII light curves, we performed time-delay analyses to find out if robust time delays can be detected for the new UV FeII templates. In particular, in comparison with the previous time-delay analyses of HE 0413-4031, here we focus on the role of the FeII emission model (template) in affecting the line-emission correlation with the continuum, and thus the time-delay determination.

 We used different time-delay methodologies, namely interpolated cross-correlation function \citep[ICCF; ][]{Peterson1998}, discrete correlation function \citep[DCF; ][]{Edelson1988}, $z$-transferred DCF \citep[zDCF; ][]{1997ASSL..218..163A}, $\chi^2$ method \citep{Czerny2013}, JAVELIN \citep{Zu2011,Zu2013,2016ApJ...819..122Z}, von Neumann, and Bartels estimators \citep{2017ApJ...844..146C} to infer the best time delay. For completeness, we show the plots and the inferred time-delay estimates for all the methods in Appendix~\ref{sec_appendix_time_delay}. For comparison, two comparably good FeII templates were considered to infer the MgII and the FeII emission time delays: the KDP15 template and the VW01 template. The uncertainties were typically inferred using the time-delay peak distributions constructed from several hundred light curves generated using the bootstrap method \citep[see e.g.,][where we already applied all the methods used in this paper]{2019AN....340..577Z,Zajacek2020,prince2023}.
 
 The time delay results for MgII emission are summarized in Table~\ref{tab_mgII}, where the values are expressed in the observer's frame unless stated otherwise. The time delays are generally between $\sim 500$ and $\sim 600$ days, except for the Bartels method and the VW01 template, which indicates the best time delay at $\sim 405$ days. The time-delay calculations for the MgII line with the new redshift give very similar results to what was obtained in \citet{prince2023} as well as in \citet{Zajacek2020} using the old redshift.  
This is related to the fact the decomposition in both cases (the older and the new redshift) yielded very similar values of the equivalent width of MgII (see lower panel of Fig.~\ref{fig:new_trends1}). 

The MgII light curve is very strongly correlated with the continuum light curve, which implies that the observed continuum is the driving photoionizing emission. In particular, the case of the best-fit FeII template KDP15 exhibits the peak correlation coefficient of $r\sim 0.87$. For the VW01 template, the correlation coefficient only slightly drops to $r\sim 0.73$. This trend is also reflected for other time-delay methods, for which the cases with the VW01 template exhibit lower correlation coefficients (zDCF) or the minima are located at larger values of the corresponding measures ($\chi^2$, von Neumann, and Bartels estimators, see Appendix~\ref{sec_appendix_time_delay}). This can be interpreted as the result of the VW01 FeII emission model, which is older and more approximate than the KDP15 model, which lowers the overall correlation of MgII with the continuum. Despite that the determined best MgII time delays are consistent within the uncertainties among the different time-delay methods and the two templates, see Table~\ref{tab_mgII}. By averaging the time delays among different methods and the two templates, we arrive at the final value of $\tau_{\rm MgII}=224^{+21}_{-23}$ days in the rest frame of the source. This value is within $2\sigma$ uncertainty consistent with the values reported in \citet{prince2023} and \citet{Zajacek2020}, whose mean values are $\sim 26\%$ and $\sim 35\%$ larger than the one found here for the new redshift and KDP15/VW01 templates.

\begin{table}[h!]
    \centering
     \caption{Summary of MgII emission time delays inferred using different methodologies. For individual methods, time delays are expressed in days in the observer's frame, while the last two rows represent the mean time delays expressed in days in the rest frame of the source.}
    \begin{tabular}{c|c|c}
    \hline
    \hline
    Method   & KDP15 & VW01  \\
    \hline
    ICCF - highest $r$ & 541 ($r=0.87$) & 606 ($r=0.73$)  \\
    ICCF - peak     & $552^{+79}_{-70}$   &  $605^{+30}_{-219}$ \\
    ICCF - centroid & $491^{+50}_{-47}$    &   $517^{+95}_{-90}$ \\
    DCF             & $565^{+34}_{-34}$   &  $586^{+29}_{-29}$  \\
    zDCF            & $597^{+35}_{-75}$   &  $638^{+49}_{-93}$ \\
    $\chi^2$        & $619^{+26}_{-116}$    &  $623^{+27}_{-62}$ \\
    Javelin         & $493^{+4}_{-15}$   &  $493^{+3}_{-17}$ \\
    Von Neumann     & $495^{+12}_{-113}$   &  $405^{+42}_{-27}$ \\
    Bartels         & $495^{+19}_{-113}$   &  $405^{+45}_{-34}$  \\
    \hline
    $\overline{\tau}$ -- observer's frame & $538^{+51}_{-56}$    & $534^{+89}_{-94}$     \\
     $\overline{\tau}$ -- rest frame    & $225^{+21}_{-23}$    &  $223^{+37}_{-39}$    \\
       \hline
    Final MgII time delay & \multicolumn{2}{c}{$224^{+21}_{-23}$} \\
    \hline
    \end{tabular}   
    \label{tab_mgII}
\end{table}

\begin{figure*}
    \includegraphics[width=\columnwidth]{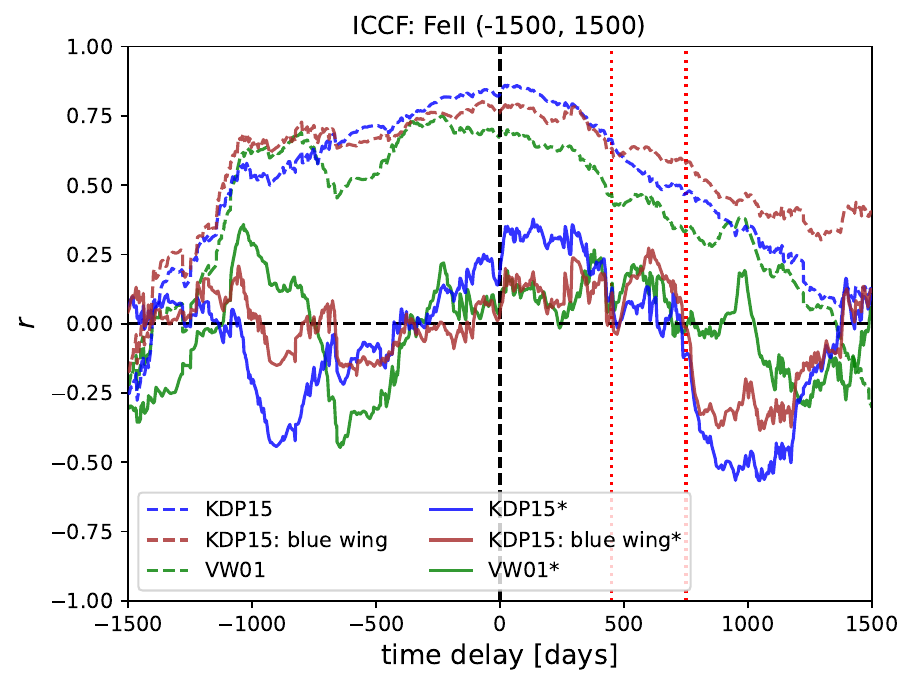}
    \includegraphics[width=\columnwidth]{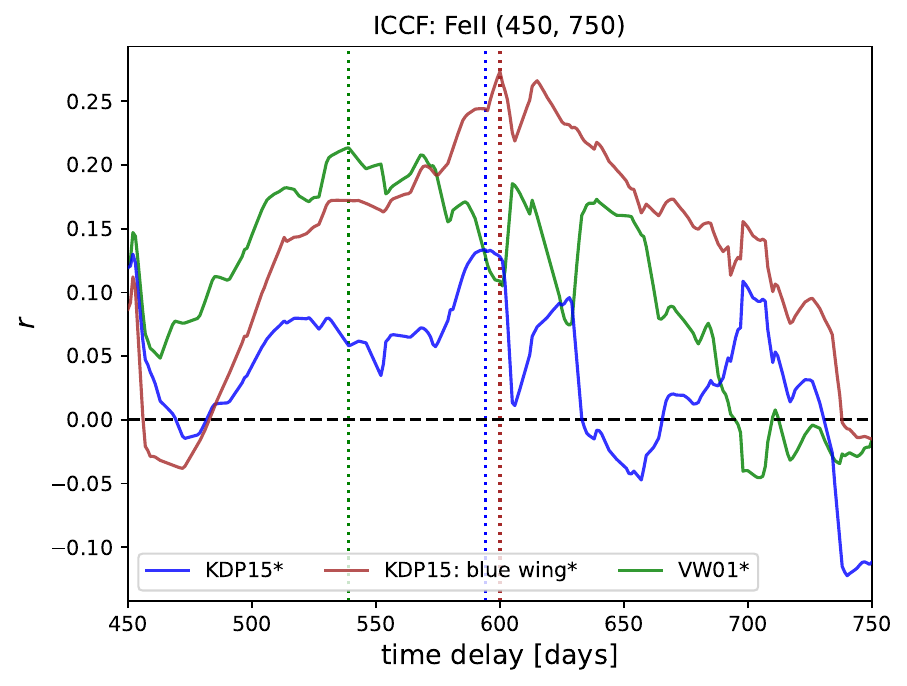}
    \caption{ICCF for the FeII emission using different FeII templates (KDP15 and VW01). \textit{Left panel:} The ICCF for the FeII templates KDP15, KDP15--blue wing, and VW01 without the long-term subtraction (dashed lines). By subtracting the long-term trend from the continuum and line-emission light curves, the ICCF becomes asymmetric with respect to zero (solid lines) and the time-delay peak between 450 and 750 days can be identified, especially for the blue-wing part of the KDP15 template. \textit{Right panel:} The ICCF for all the three considered cases (KDP15, KDP15--blue wing, VW01) in the interval between 450 and 750 days in the observer's frame. For the best-fitting KDP15 template, the cross-correlation peak is at $\sim 600$ days. See Table~\ref{tab_feII} for the summary of the peak and the centroid values (last three rows).}
    \label{fig_iccf_feII}
\end{figure*}

\begin{table*}[h!]
    \centering
     \caption{Summary of FeII emission time-delay constraints (expressed in days in the observer's frame) using the ICCF. The results denoted with the star (ICCF$^{\star}$) represent the cases with the long-term trend subtraction (using the third-order polynomial fit) from both light curves. The time interval in the parentheses denotes the search interval for the best time delay.}
    \begin{tabular}{c|c|c|c}
    \hline
    \hline
    Method   & KDP15 & KDP15 - blue &  VW01  \\
    \hline
    ICCF - highest  $r$  (-1500,+1500) & 50 ($r=0.86$)   & -71 ($r=0.80$)   &  -237 ($r=0.75$) \\
    ICCF - peak (-1500,+1500)       & $34^{+62}_{-132}$   & $22^{+187}_{-542}$   &  $-237^{+270}_{-122}$ \\
    ICCF - centroid (-1500,+1500)    & $-38^{+93}_{-109}$   & $-69^{+165}_{-269}$   & $-85^{+92}_{-215}$ \\
    \hline
    ICCF$^{\star}$ - highest $r$ (-1500, +1500) & 134 ($r=0.38$)   & 600 ($r=0.27$)   & $-1034$ ($r=0.36$) \\
    ICCF$^{\star}$ - peak (-1500, +1500)        & $134^{+162}_{-234}$   & $128^{+472}_{-1194}$   &  $-31^{+569}_{-1016}$ \\
    ICCF$^{\star}$ - centroid (-1500, +1500)    & $127^{+183}_{-219}$   & $114^{+483}_{-1184}$   & $-50^{+580}_{-990}$ \\
    \hline
    ICCF$^{\star}$ - highest $r$ (450, 750) & 594  ($r=0.13$)   &  600 ($r=0.27$)   & 539 ($r=0.21$) \\
    ICCF$^{\star}$ - peak (450, 750)   &  $595^{+96}_{-66}$   & $600^{+31}_{-10}$   &  $574^{+67}_{-43}$ \\
    ICCF$^{\star}$ - centroid (450, 750)    & $592^{+102}_{-55}$ & $602^{+28}_{-30}$   & $570^{+76}_{-39}$ \\
    \hline
    Final FeII time delay (rest frame) & \multicolumn{3}{c}{$251^{+9}_{-7}$}\\
    \hline
    \end{tabular}   
    \label{tab_feII}
\end{table*}

\begin{table}[]
    \centering
     \caption{Comparison of correlation coefficients for different FeII templates for time delays below and above zero days (up to $\mp 800$ days). After the long-term trend subtraction from both continuum and FeII light curves, the ICCF becomes clearly asymmetric, and for time delays larger than zero days, the mean correlation coefficient is greater and positive. The templates denoted by stars represent the light curves with the long-term trend subtraction (last three rows).}
    \begin{tabular}{c|c|c}
    \hline
    \hline
    Template   & (-800, 0) & (0, 800)  \\
    \hline
    KDP15                 & 0.74    & 0.68    \\
    KDP15--blue-wing      & 0.71    & 0.68    \\ 
    VW01                  & 0.63    & 0.53    \\
    \hline
    KDP15*                 & -0.01    & 0.16    \\
    KDP15*--blue-wing      & -0.04    & 0.11    \\ 
    VW01*                  & -0.08    & 0.10    \\
    \hline
    \end{tabular}   
    \label{tab_correlation_asymmetry}
\end{table}

\begin{figure*}
    \includegraphics[width=\columnwidth]{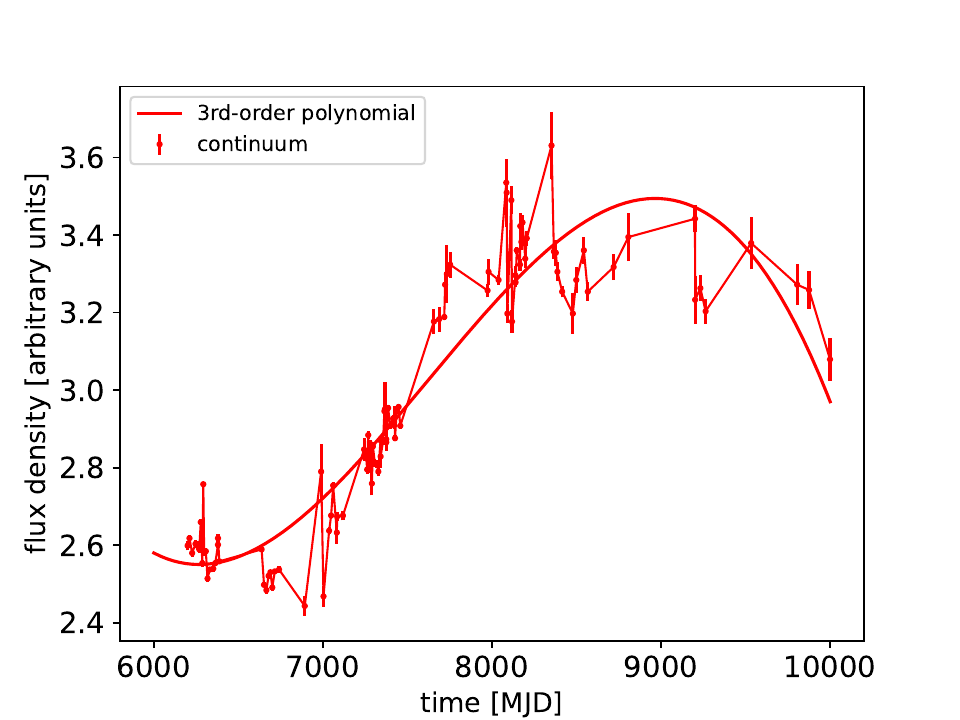}
    \includegraphics[width=\columnwidth]{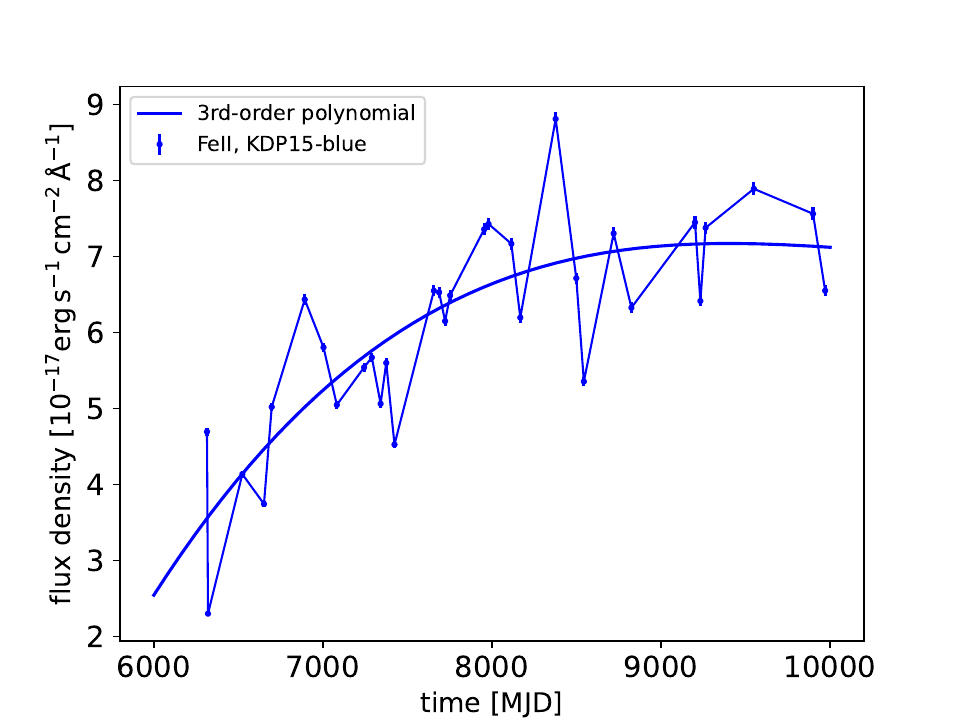}
    \includegraphics[width=\columnwidth]{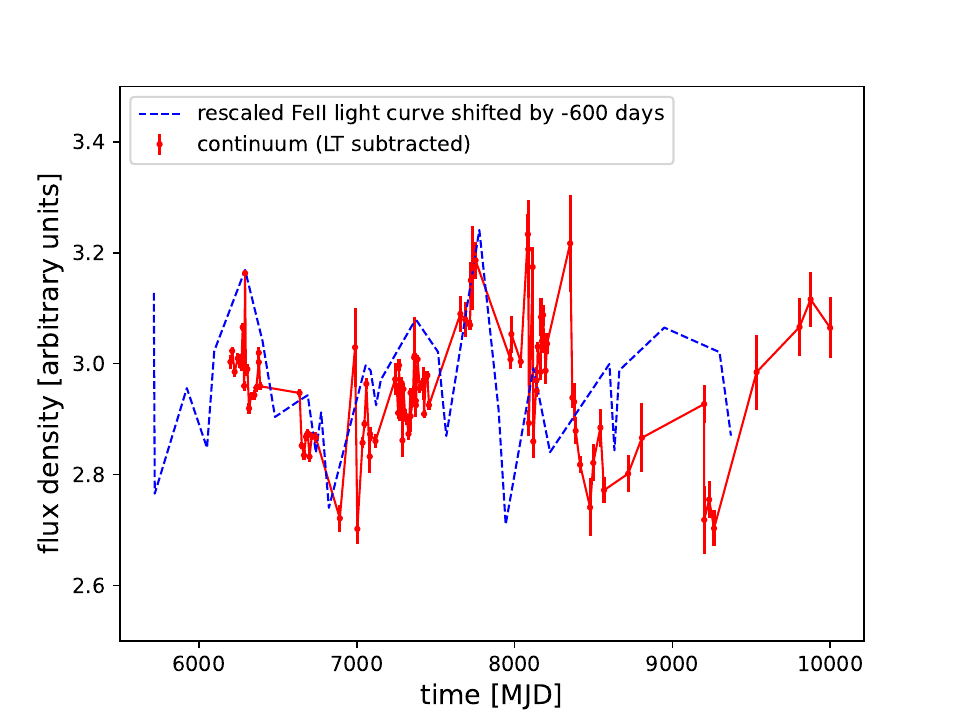}
    \includegraphics[width=\columnwidth]{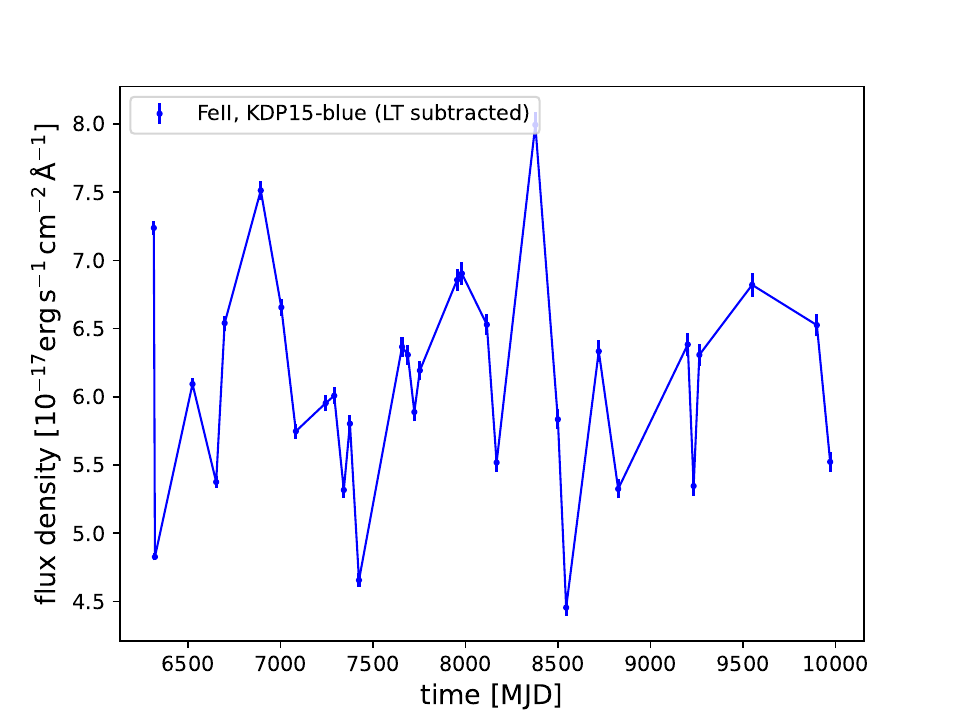}
    \caption{Time-delay analysis using the long-term (hereafter LT) trend subtraction from continuum and FeII emission light curves. \textit{Upper panels:} Continuum (left) and FeII emission light curve (right), which was inferred using the KDP15 template considering the more prominent blue wing, with the corresponding 3rd-order polynomial fits. \textit{Lower panels:} Continuum and FeII emission light curves in the left and the right panels, respectively, after the LT trend was subtracted from both of them. The final light curves were reconstructed by adding the mean flux density value of the corresponding original light curves. In the left panel for the continuum light curve, we also show the rescaled FeII light curve shifted by -600 days, which corresponds to the ICCF maximum for this case.}
    \label{fig_lc_subtract}
\end{figure*}

However, the FeII intensity is much different in these two cases. The FeII emission in the new KDP15 decomposition based on the new redshift is now by a factor of 3 fainter. We performed time-delay computations but we did not obtain a satisfactory solution for the time delay. The situation is best illustrated in the case of the ICCF plot (see Figure~\ref{fig_iccf_feII}, left panel, dashed lines). The maximum value of the correlation coefficient $r$ is close to 0.8 for both KDP15 and VW01 templates which clearly shows that the FeII light curve and the continuum are very well correlated. However, the function peaks close to the zero time delay, see the summary of peak and centroid values for both templates in Table~\ref{tab_feII} (first three rows). This strongly suggests that the FeII light curve is strongly contaminated by the continuum itself and the decomposition of the spectrum into FeII and continuum is probably not satisfactory. 

In the next attempt to search for a FeII time delay, we just consider the blue wing of the FeII emission in the wavelength range 2683.09-2800.00\,\AA, which is dominated by FeII multiplet 63. The blue wing of the FeII emission is more pronounced within the KDP15 model, see Figs.~\ref{fig:new_redshift} and \ref{fig:Fe_aver}, and thus less contaminated by the continuum. This does not result in a qualitative change of the ICCF and the peak correlation coefficient is again close to zero, see Figure~\ref{fig_iccf_feII} and Table~\ref{tab_feII}.

Finally, we assess that the large peak close to zero days is due to the continuum long-term trend present in both light curves. For the FeII emission light curve, the trend is present due to significant contamination by continuum. Therefore, to isolate the FeII variability and response, we subtract the long-term (hereafter LT) trend from both light curves using the fitted third-order polynomial in the form $f_{\rm trend}=f_1(t-t_0)^3+f_2(t-t_0)^2+f_3(t-t_0)+f_4$, where $f_1$, $f_2$, $f_3$, and $f_4$ are the fitted parameters. We show the continuum and the UV FeII emission light curves after the LT trend subtraction in Figure~\ref{fig_lc_subtract} for the case of the KDP15 template (the FeII emission corresponds to the more prominent blue wing), wherein the top panels we display the original light curves and in the bottom panels we show the corresponding light curves after the LT term subtraction and with the addition of the mean flux densities of the original light curves. This proves efficient in enhancing the correlated and delayed FeII emission (see Figure~\ref{fig_iccf_feII}, solid lines), which is revealed by the asymmetric ICCF that is increasing for positive time delays. When we consider negative and positive time delays in the ranges (-800, 0) days and (0,+800) days, the calculated mean values of correlation coefficients for the templates KDP15, KDP15--blue wing, and VW01 are listed in Table~\ref{tab_correlation_asymmetry}, where we see that after the LT trend subtraction (templates denoted by the star), the ICCF gets asymmetric with the mean correlation coefficients for time delays above zero days being greater and positive. For the large time-delay interval of (-1500, 1500) days, the peak and centroid values become larger and mostly positive with respect to the cases without the LT trend subtraction, see Table~\ref{tab_feII}. However, this is at the cost of generally lower correlation with the peak values of only $r\sim 0.3-0.4$. The well-defined correlation peak at $\sim$ 600 days (in the observer's frame) for the KDP15 FeII template--blue wing is considered as the best candidate time-delay peak. Thus, when we narrow down the search interval to (450, 750) days for KDP15, KDP15--blue wing, and VW01 templates, all three cases share consistent peak and centroid ICCF values close to $\sim 600$ days, see Table~\ref{tab_feII} (last three rows). This is also confirmed visually in Figure~\ref{fig_lc_subtract} (bottom left panel), where we superpose the rescaled FeII light curve shifted by -600 days. Although these time-delay values correspond to rather low correlation coefficients, we treat the time delay of $\tau_{\rm FeII}=251^{+9}_{-7}$ days (average of the peak and the centroid values for the KDP15-blue wing case expressed in the rest frame) as a plausible candidate value of the true FeII emission time delay. 

In Appendix~\ref{subsec_sub}, we also verify that after the subtraction of the LT trend from the MgII light curve, it is possible to recover the time delay reported previously. Using the ICCF, we can recover a time-delay peak between $\sim 400-600$ days (in the observer's frame), albeit at a significantly lower correlation coefficient of $\sim 0.4-0.5$. When the centroid and the peak values are averaged among the two templates, the mean rest-frame time delay $213^{+62}_{-35}$ days is in agreement with the previously reported time delay without any LT trend subtraction. Hence, in summary, the LT trend subtraction can help recover a time delay between the two light curves, especially for line light curves contaminated by the underlying continuum emission, such as for the FeII emission in this paper, although it is at a cost of lowering the correlation coefficient and increasing the uncertainty in the time-delay determination. 

%The blue wind is stronger, and the red is much fainter, so in order to

\section{Discussion}
\label{sec:discussions}

We analyzed the spectroscopic data collected in the years 2013 - 2023 with the telescope SALT for the intermediate-redshift quasar HE~0413-4031, concentrating on the MgII spectral region. These data include one more spectrum than the data analyzed by \citet{prince2023} but most importantly, in the meantime, we determined the precise redshift to the quasar from the IR observation at the observed frame with the telescope SOAR which is based on the narrow optical [OIII] $\lambda \lambda 4959,5007$ lines considered as the most reliable indicators of the quasar rest frame. New redshift determination essentially changed the decomposition of the quasar UV spectrum into MgII, FeII, and the underlying continuum. We checked how it affected the time delay measurements.

We showed that the MgII time delay was not affected by the change of the redshift and decomposition. The new decomposition with the UV FeII KDP15 template implies no shift with respect to the systemic redshift while the previous redshift/decomposition indicated the velocity shift between the FeII and MgII. Nevertheless, the time delay remained the same, from 500 to 600 days in the observed frame, independent of the decomposition, redshift, and the time delay measurement method. Thus the maximum systematic error is on the order of 10 \%. We thus stress that the MgII line time delay can be used reliably for probing the properties of the BLR and for cosmology, and the results are not sensitive to the actual method.

However, the line properties of the FeII emission are very sensitive to the redshift and the best-fitting FeII model. Previously, in \citet{prince2023} we performed the time-delay analysis of the total FeII emission as well as wavelength-resolved reverberation mapping of the MgII+FeII complex. The recovered MgII and FeII time delays based on different methods, $314.4$ and $330.6$ days, respectively, are in tension with the MgII and FeII FWHMs, FWHM(MgII)$=4380\,{\rm km\,s^{-1}}$ and FWHM(FeII)$=2820\,{\rm km\,s^{-1}}$, respectively. Assuming that the MgII and FeII material is fully virialized with the same virial factor, the FeII time delay should be significantly larger than the one reported by \citet{prince2023}, $\tau_{\rm FeII}=314.4(4380/2820)^2\,\text{days}\sim 758\,\text{days}$. In the current analysis, with the new redshift and the new KDP15 template, the inferred MgII and FeII time delays and the corresponding FWHMs appear to be fully in agreement since the FeII rest-frame time delay expected from the virialized motion, $\tau_{\rm FeII}=\tau_{\rm MgII} (\text{FWHM(MgII)}/\text{FWHM(FeII)})^2\sim 224(4503.1/4200)^2\,\text{days}\sim 257\,\text{days}$, is consistent with the FeII time delay inferred from reverberation, $\tau_{\rm FeII}=251^{+9}_{-7}$ days. This implies that FeII emitting material is more distant than MgII line-emitting material by $\sim 27$ light days or 0.023 pc ($\sim 4700\,{\rm AU}$).  

The quasar HE 0413-4031 does not seem to exhibit a separate very broad Gaussian component forming the wings of the MgII line as it is typical for lower Eddington sources \citep[quasar type B population;][]{Sulentic2000,KDP2015}, so the line is well modeled by a single Lorentzian shape which helps to disentangle better FeII and MgII emission. This makes the decomposition into MgII and FeII relatively easier than in lower Eddington sources. This property remained true for all our experiments with the two redshifts and various FeII templates.  HE 0413-4031 is therefore a good candidate for further FeII studies, complementing the nearby object I Zw 1.

An even more advanced technique is to use the wavelength-resolved time delays which allow us to trace separately the wings of the lines from the core as well as the kinematics of the BLR medium when multiple lines contribute to a given wavelength range. Such studies were done for more than 35 AGNs by various authors (\citealt{Bentz2010}; \citealt{Denney2010}; \citealt{Grier2012}; \citealt{Du2018}; \citealt{derosa2018}; \citealt{Xiao_2018}; \citealt{Zhang2019}; \citealt{Hu_2020}; \citealt{U2021}) for low-redshift sources. When we analyzed the SALT data for the quasar HE 0413-4031 by subtracting the underlying continuum but without separating the MgII and FeII, we concluded that the FeII emission originates in a region slightly more distant than MgII \citep{prince2023}. So the issue of the FeII emission region geometry is much more complex due to the pseudo-continuum character of this component than MgII which gives firm and reliable results.

The UV FeII emission template is also relevant for estimating the FeII pseudocontinuum and MgII emission contribution to the AGN UV emission in near-UV photometric bands. This is especially relevant for the future monitoring of nearby AGN using one-band \citep[e.g., the \textit{ULTRASAT} mission;][]{2023arXiv230414482S} or two-band UV photometry \citep[e.g., the \textit{QUVIK} mission;][]{2022SPIE12181E..0BW,2023arXiv230615080W,2023arXiv230615081K}, with the attempt to perform UV continuum reverberation mapping of accretion disks \citep{2023arXiv230615082Z}. The reprocessing of the harder radiation by the BLR clouds dilutes the disk temporal response, which affects the continuum time delay measurements in a distinct way \citep{2022MNRAS.509.2637N}. The correct understanding of the UV FeII multiplet emission and its geometric distribution will help to account for and understand the reprocessing contribution by the BLR gas.  

\subsection{HE0413 and UV MgII \& FeII $R-L$ relations}

\begin{figure*}
    \includegraphics[width=\columnwidth]{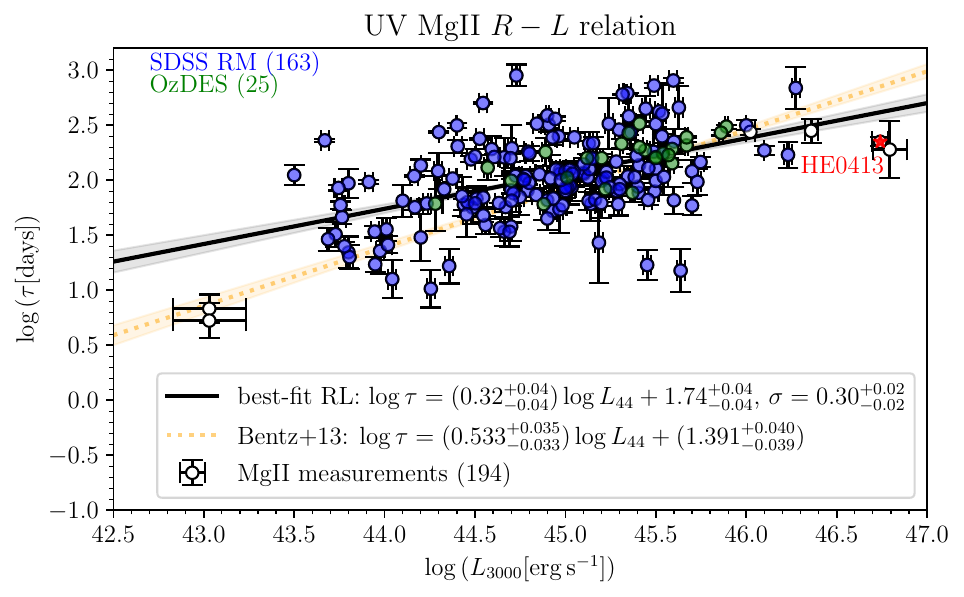}
    \includegraphics[width=\columnwidth]{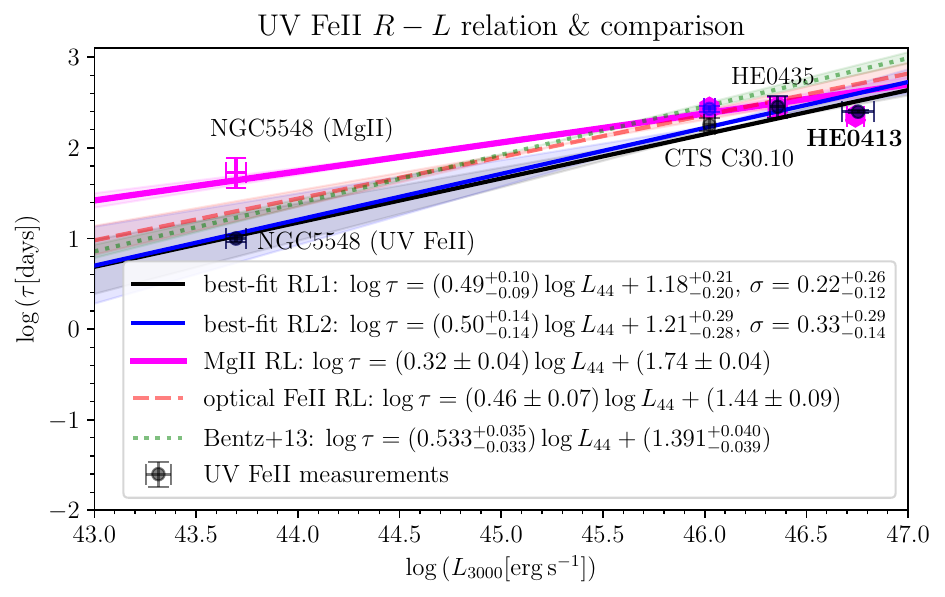}
    \includegraphics[width=\columnwidth]{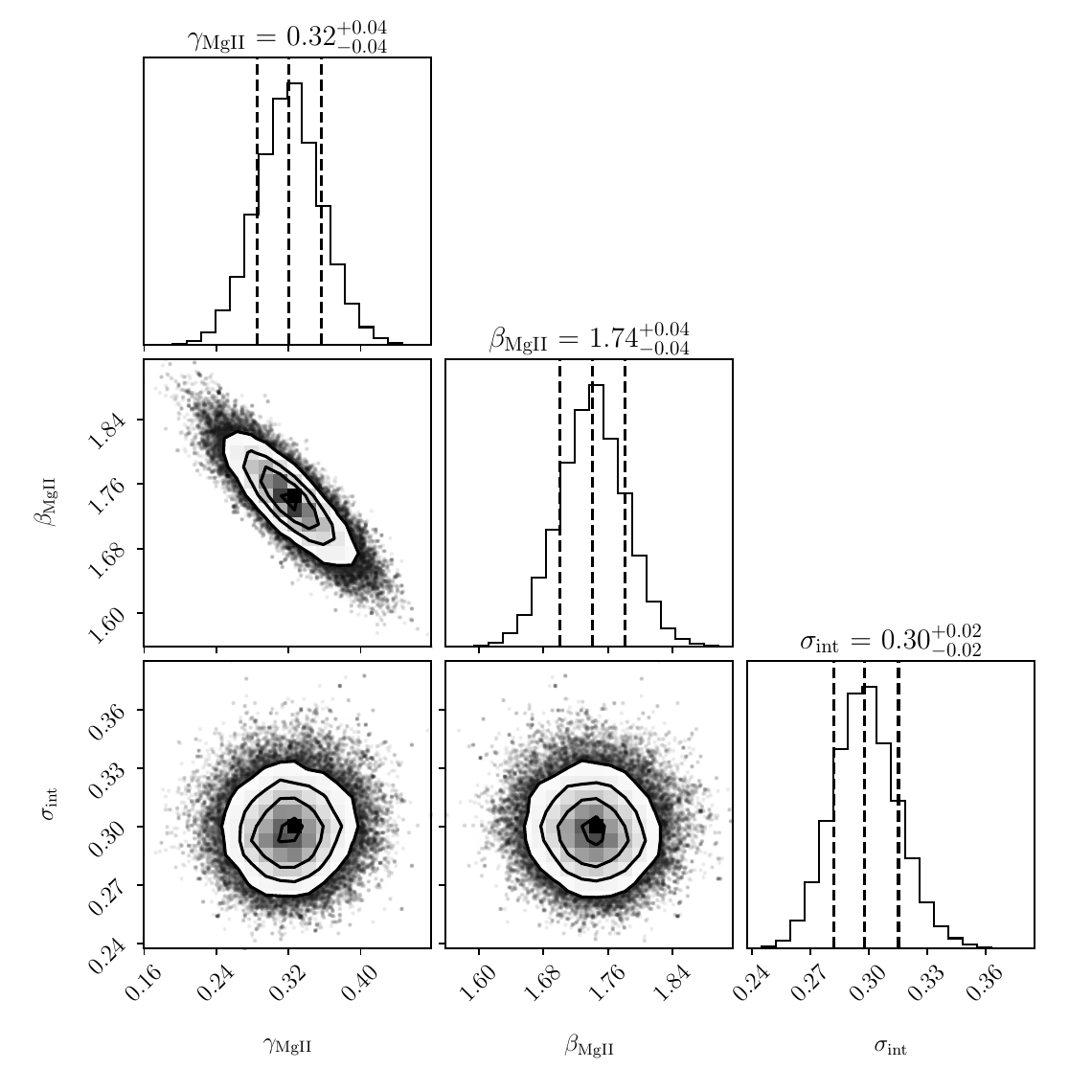}
    \includegraphics[width=\columnwidth]{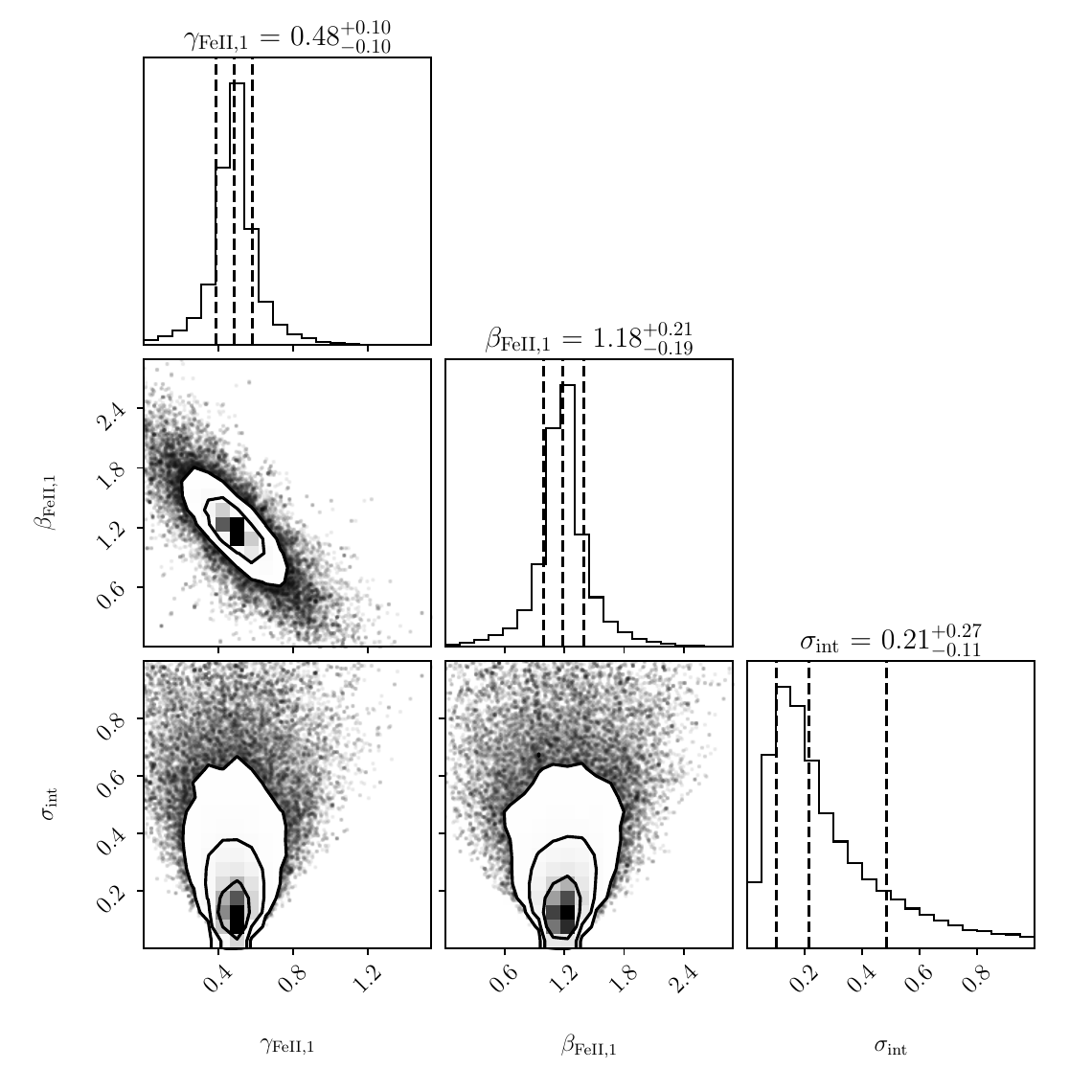}
    \caption{Position of HE 0413-4031 in the UV MgII and FeII $R-L$ relations. \textit{Upper panels:} MgII $R-L$ relation (left panel) constructed from 194 measurements (163 SDSS RM, 25 OzDES, and six other sources including HE 0413-4031). The best-fitted relation is in the legend. For comparison, we also include the H$\beta$ relation \citep{Bentz2013} recalculated for 3000\,\AA\, (dashed orange line). In the right panel, we show the UV FeII $R-L$ relation based on four RM measurements for NGC 5548, CTS C30.10, HE 0435-4312, and HE 0413-4031 (two comparable relations are plotted by solid black and dashed blue lines since there were two plausible FeII time delays for CTS C30.10). The UV FeII relation is steeper and lower than the flatter and higher MgII $R-L$ relation, which results in the convergence and their intersection towards higher luminosity sources. In the right panel, we also show for comparison the optical FeII relation inferred by \citet{prince2023} from multiple measurements as well as the H$\beta$ $R-L$ relation by \citet{Bentz2013} that are depicted by dashed red and green lines, respectively.  \textit{Lower panels:} One-dimensional confidence histograms and two-dimensional confidence contours for the searched parameters (slope $\gamma$, intercept $\beta$, and the intrinsic scatter $\sigma_{\rm int}$) of the UV MgII $R-L$ relation (left panel) and of the UV FeII $R-L$ relation (right panel). For the UV FeII $R-L$ relation, we show the case with a smaller scatter, and hence the smaller FeII time delay of CTS C30.10.  }
    \label{fig_MgII_FeII_RL}
\end{figure*}

Given the new redshift and best-fit FeII template, which resulted in slightly modified MgII and FeII time delays, we construct an updated UV MgII radius-luminosity relation or $R-L$ relation. We build upon its previous versions presented in \citet{Czerny2019}, \citet{Zajacek2020}, \citet{2020ApJ...903...86M}, \citet{zajacek2021}, and \citet{prince2023}, where we included a progressively larger number of MgII RM quasars \citep[see also][]{Homayouni2020,2023MNRAS.522.4132Y}.

Here we combine the sample of RM quasars with significant ($>2\sigma$) detections from the SDSS RM program \citep[163 sources; ][]{Shen2016,Homayouni2020,2023arXiv230501014S}, OzDES RM program \citep[25 sources; ][]{2023MNRAS.522.4132Y}, and 6 sources from other RM campaigns \citep{Metzroth2006,Lira2018,Czerny2019,2022A&A...667A..42P,prince2023}. Altogether we collected 194 measurements, which is one of the largest samples of RM MgII AGN up to today. The sample exhibits a significant positive correlation between the rest-frame time delays and monochromatic luminosities at 3000\,\AA\, with the Pearson correlation coefficient of $0.54$ ($p=2.22 \times 10^{-16}$) and the Spearman rank-order correlation coefficient of $0.50$ ($p=8.56 \times 10^{-14}$). Next we fit the power-law relation $\log{(\tau\,[\text{days}])}=\beta+\gamma\log{(L_{3000}/10^{44}\,{\rm erg\,s^{-1}})}$ to the data. Using the Markov Chain Monte Carlo (MCMC) approach implemented in the python package \texttt{emcee}, we infer the best-fit MgII $R-L$ relation,
\begin{equation}
    \log{\left(\frac{\tau}{\text{day}}\right)}=(1.74^{+0.04}_{-0.04})+(0.32^{+0.04}_{-0.04})\log{\left(\frac{L_{3000}}{10^{44}\,{\rm erg\,s^{-1}}} \right)}\,.
    \label{eq_MgII_RL}
\end{equation}
We show the MgII $R-L$ relation including the best-fit relation given by Eq.~\eqref{eq_MgII_RL} in Figure~\ref{fig_MgII_FeII_RL}, see the top left panel. In the bottom left panel of Figure~\ref{fig_MgII_FeII_RL}, we display one-dimensional likelihood distributions and two-dimensional likelihood contours of the searched parameters. We can especially notice a distinct degeneracy (negative correlation) between the intercept and the slope. However, despite this degeneracy, the inferred slope is much flatter than 0.5, which is expected from arguments based on the ionization parameter (\citealt{davidson1979}; see also the discussion in \citealt{Bentz2013}) or from the dust-based BLR model \citep{czhr2011}. The inferred intrinsic scatter is substantial, $\sigma_{\rm int}=0.30^{+0.02}_{-0.02}$, however, it is moderate given a relatively large number of time-delay measurements and comparable to the scatter found previously for a significantly smaller number of sources \citep{prince2023,Homayouni2020}. 

We also revisit the UV FeII $R-L$ relation, which was first presented in \citet{2022A&A...667A..42P} for 2 measurements only and in \citet{prince2023} for 4 measurements, given the updated time delay for HE0413-4031 presented in Subsection~\ref{subsec_time_delay}. The two best-fit relations (RL1 and RL2) and their corresponding parameters found using the same MCMC approach as for the MgII are shown in Figure~\ref{fig_MgII_FeII_RL} (top and bottom right panels). The two relations stem from the two plausible values for the UV FeII time delay for the quasar CTS C30.10 as found by \citet{2022A&A...667A..42P}. This difference is, however, small and the two relations are identical within the uncertainties. The smaller RMS scatter is exhibited by the first UV FeII $R-L$ relation (RL1 in Fig.~\ref{fig_MgII_FeII_RL}) where we take into account the smaller UV FeII time delay of CTS C30.10 ($\tau_{\rm FeII}= 180.3^{+26.6}_{-30.0}$ days). The best-fit relation is, 
\begin{equation}
    \log{\left(\frac{\tau}{\text{day}}\right)}=(1.18^{+0.21}_{-0.20})+(0.49^{+0.10}_{-0.09})\log{\left(\frac{L_{3000}}{10^{44}\,{\rm erg\,s^{-1}}} \right)}\,,
    \label{eq_MgII_RL}
\end{equation}
with the intrinsic scatter of $\sigma_{\rm int}=0.22^{+0.26}_{-0.12}$. For this relation, we show the one-dimensional likelihood distributions and two-dimensional likelihood contours in Figure~\ref{fig_MgII_FeII_RL} (bottom right panel). 

Although the number of UV FeII time-delay measurements is limited, we see in Figure~\ref{fig_MgII_FeII_RL} that there is a tendency for UV FeII $R-L$ relation to be positioned below the MgII $R-L$ relation because of the smaller intercept. However, it appears to be steeper than the MgII $R-L$ relation, with the slope $\gamma\sim 0.5$ consistent with the standard photoionization theory of the BLR clouds. This results in larger differences between MgII and UV FeII time delays for lower luminosity sources, for which $\tau_{\rm MgII}>\tau_{\rm FeII}$, while at higher luminosities, both relations converge and intersect, which results in $\tau_{\rm MgII}\lesssim \tau_{\rm FeII}$. This also appears to be the case of HE 0413-4031 which is effectively at the high luminosity end of both MgII and UV FeII $R-L$ relations. This trend also seems to be in contrast with the $R-L$ relations in the optical domain, where both H$\beta$ \citep{Bentz2013} and optical FeII $R-L$ \citep{prince2023} relations are nearly identical with each other within the uncertainties when expressed for 3000\,\AA, see Figure~\ref{fig_MgII_FeII_RL}. Their slopes are also in agreement with the simple photoionization arguments unlike MgII $R-L$ relation, which is significantly flatter.     

\subsection{Simple tests of \texttt{CLOUDY} models of the FeII emission}

 The three FeII templates we use are almost comparable in the fit quality for Observation 14 (see Figure~\ref{fig:new_redshift}) but their shapes are considerably different. To have some insight into the physical nature of each template, we use the recent version C22.01 of the code \texttt{CLOUDY} to model the reprocessed continuum \citep[see also][]{pandey2023}. As in \citet{Bruhweiler2008}, we parameterize the \texttt{CLOUDY} solution with the incident flux, $\log \phi$, the local density of the cloud, $\log n_{\rm H}$, and the turbulent velocity, $v_{\rm turb}$. The results are discussed below.

\begin{figure}
    \centering
    \includegraphics[width=\columnwidth]{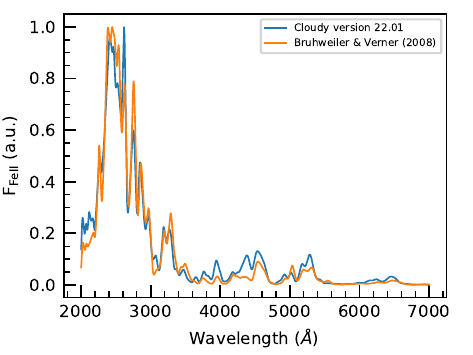}
    \includegraphics[width=\columnwidth]{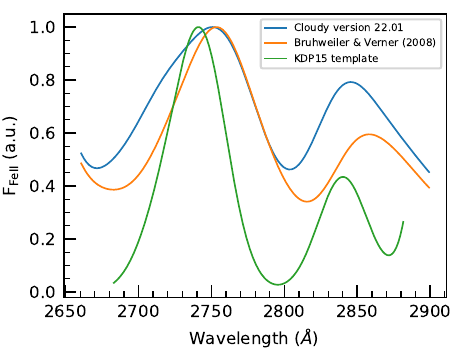}
    \includegraphics[width=\columnwidth]{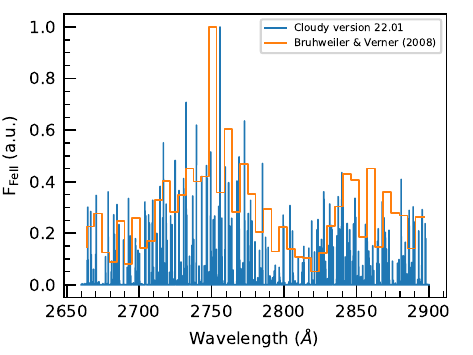}
    \caption{Comparison of \citet{Bruhweiler2008} template with the \texttt{CLOUDY} C22.01 output for $\log{(\phi [{\rm cm^{-2} s^{-1}}]} =20.5$, $\log{n_{\rm H} [{\rm cm^{-3}}]}=12$, the turbulent velocity of 20 km/s, and the column density of 10$^{24}$ cm$^{-2}$. Upper panel: For a wavelength range of 2000-7000 $\AA$ with a Gaussian broadening assuming FWHM $\sim$4200 km/s. Middle panel: The zoomed-in version of the upper panel in the wavelength range 2650-2900 \AA. Here we also add the broadened KDP15 template for comparison (green solid line). Lower panel: The nonbroadened version of the middle panel. Here, the templates are renormalized for a better comparison.}
    \label{fig:d12_and_new}
    %For new Cloudy calculations log $\phi$=20.5, log n$_H$=12, turbulence velocity=20 km/s, column density of 10$^{24}\,{\rm cm^{-2}}$, with broadening by a Gaussian assuming FWHM $\sim$4200 km/s (upper panel). The middle panel shows a zoomed-in version within the wavelength range 2650-2900 \AA. Here, the templates are re-normalized for better comparison. The lower panel shows non-broadened data as taken from \citet{Bruhweiler2008} and CLOUDY version 22.01.  
\end{figure}

 The template d12 used in \citet{prince2023} was obtained with the \texttt{CLOUDY} several years ago, and in the meantime, the atomic content of the code improved significantly. We thus compared the available template d12 with the output of the \texttt{CLOUDY} (version 22.01) using the same input parameters: $\log{(\phi [{\rm cm^{-2} s^{-1}}])} =20.5$, $\log{(n_{\rm H} [{\rm cm^{-3}}])} =12$, and the turbulent velocity of 20 km/s. The column density was assumed to be $10^{24}$ cm$^{-2}$. The incident SED was very similar, that is a standard AGN SED in our computation, and the SED from \cite{1997ApJS..108..401K} in \citet{Bruhweiler2008}. We smeared the results with the Gaussian profile corresponding to the FWHM of 4200 km/s as used in the current paper. The upper panel of Figure~\ref{fig:d12_and_new} shows the FeII emission over the broadband range (2000-7000 \AA). %with a very considerable decrease in the UV part in comparison to the optical part
 In the middle panel, we show just the enlarged UV part, relevant to the current work. It shows that, compared to the \citet{Bruhweiler2008} template, there is a marginal shift in the peak of FeII pattern towards a shorter wavelength in the case of the new \texttt{CLOUDY} output; more importantly, though, the bluer portion of the FeII pattern is more asymmetric.
 Thus, this template is more consistent with the new redshift when we look at the position of the blue wing of the VW01 and KDP15 templates (see also Figure~\ref{fig:new_redshift}). Comparing the red-wing peak positions, we see that the red peak of the VW01 template (or d12) is shifted towards the red part and that of KDP15 is shifted slightly towards blue or lower wavelengths. In terms of the intensity, VW01 template is overpredicting the red wing while the d12 and KDP15 templates show a fainter red-wing (see Figure~\ref{fig:new_redshift}, middle panel). In addition, the maximum of the KDP15 template is close to 2740 \AA~, hence at a slightly smaller wavelength in comparison to the new \texttt{CLOUDY} template. However, the blue peaks of both the old d12 template and the KDP15 template are rather symmetric, while the new \texttt{CLOUDY} template exhibits a prominent blueward asymmetry. 

That rather implies that new templates should be constructed for a broader range of parameters and compared to the data which is beyond the scope of the current paper. The KDP15 template has 6 parameters but the setup does not have the required flexibility since finally only two of them were used efficiently, with the other four going frequently to zeros (see Section~\ref{sect:decompo} and Figure~\ref{fig:Fe_aver} for the discussion). Their transitions peak at (maybe) too long wavelengths (2740 \AA - transition 63 and 2747 \AA - transition 62) in the blue wing part to represent well the pattern. The actual number of FeII transitions is very high, as we illustrate in the lower panel of Figure~\ref{fig:d12_and_new}. The strongest transition is close to $\sim 2750$ \AA~ in both cases. However, in the new computations, some of the transitions at shorter wavelengths, close to 2730 \AA~, become stronger, pushing the smeared pattern toward the blue part.

We also checked the \texttt{CLOUDY} output for the ionization parameter $\log{(\phi\,[{\rm cm^{-2}s^{-1}}])} = 20$, local hydrogen density $\log{(n\, [{\rm cm^{-3}}])} = 12$, and the turbulent velocity 20 km $^{-1}$ to see if there are any transitions close to the two peaks visible in the empirical I Zw 1 component: 2715.20 \AA~ and 2839.88 \AA.  Indeed, we see a strong FeII transition at 2715.60 \AA (transition $^2G^{0}_{9/2}$ $\rightarrow$ $^2G_{7/2}$\footnote{The exact transition was identified using the NIST Spectral Database: \url{https://physics.nist.gov/PhysRefData/ASD/lines_form.html}}), with the intensity $4.0 \times 10^5\,{\rm erg\,s^{-1}\,cm^{-2}}$, and the transition 
at 2839.80 \AA (transition $^6D_{9/2}$ $\rightarrow$ $^6p^0_{7/2}$), with the intensity of $1.2 \times 10^6\,{\rm erg\,s^{-1}\,{\rm cm^{-2}}}$. The ratio of the second to the first one is about 3.08, as in the case of the two peaks in the I Zw 1 component in the KDP15 template (2.80).
%However, we cannot identify to which multiplet these transitions belong.

 In summary, the computations with the use of the latest \texttt{CLOUDY} version give a different shape of the FeII emission than the old d12 template. The shift between the two is seen in the middle panel of Fig.~\ref{fig:d12_and_new} where we plot the broadened version of the templates, but none perfectly represents the peak obtained with the KDP15 template. That shows that there is room for an improvement of the standard one-dimensional FeII templates (not based on several multiplets as in KDP15) as well, but the broad search of the parameter space is beyond the scope of the current paper.  It should also be noted that such shifts in the templates can partially be responsible for the large systemic shifts reported in the literature. For example, \citet{rusakov2023} reports a shift of 4100 km s$^{-1}$ of the broad H$\beta$ line with respect to the narrow component although the MgII shift is moderate, only by -1200 km s$^{-1}$, and they propose a shock wave due to the recent supernova as an explanation. But the effect might also be due to the applied FeII template, this time in the optical band.

\subsection{Quasar location in the UV plane}

The quasar optical plane is a very convenient way to study the statistical properties of quasars \citep{Sulentic2000, Shen_Ho_2014, Sulentic_Marziani_2015FrASS, Fraix-Burnet_etal_2017FrASS, Marziani_etal_2018FrASS, panda2018, panda2019}. It is based on a two-dimensional plot using the FWHM(H$\beta$) and the parameter $R_{\rm Fe} = EW(FeII_{\rm opt})/EW(H\beta$). Here optical FeII is usually measured close to H$\beta$, in the spectral range of 4434-4684\AA~ \citep{boroson1992, Marziani_etal_2018FrASS, panda2018}. The study of the quasar main sequence allows us to understand the diversity of Type-1 AGNs and how their spectral properties are tied to the fundamental black hole parameters, such as their mass and accretion rate \citep{Marziani_etal_2018FrASS,panda2018, panda2019, Panda_etal_2019_WC,Panda_2021PhDT........22P} and the physical conditions of the broad line-emitting region \citep{panda2018, panda2019, Sniegowska_etal_2021, Panda_etal_2022FrASS, Shimeles_2023MNRAS}.

When the near-UV spectra are studied, a similar concept of the UV plane can be introduced \citep{UVplane2020}. A large sample of objects was located on this plane using the QSFIT package \citep{2017MNRAS.472.4051C} for data analysis. This package is based on the VW01 FeII template. For the quasar studies in this paper, we also have the results of fitting the VW01 template for Observation 14. The difference between the QSFIT and our fitting is in the wavelength range which covers 1250 - 3090 \AA~ while we fit only the range  2683.09448 - 2881.49854. To make the values of EW comparable, we integrated the VW01 template in the two ranges and determined the scaling parameter 6.39345. We adopted the second solution for this template from Table~\ref{tab:Fe} since the broadened version works better. After rescaling, the parameters of UV FeII are the following: EW(FeII) = 88.23 \AA, and $R_{Fe} = 2.37$.

%----------------------------------------------------------------- 
   \begin{figure}
   \centering
\includegraphics[width=0.95\linewidth]{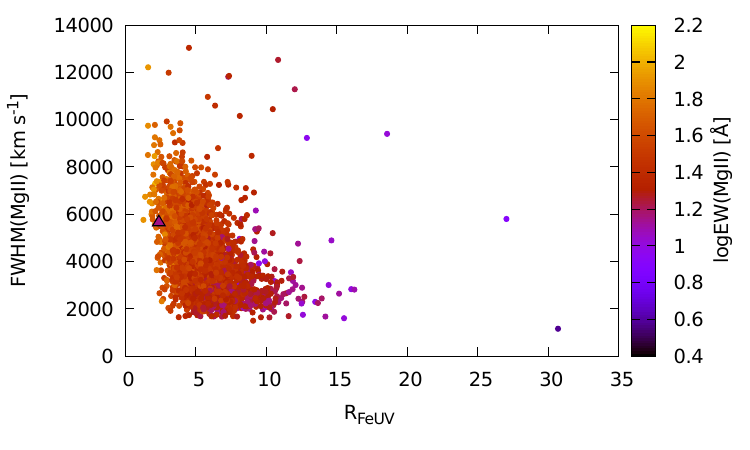} 

      \caption{Quasar UV plane with sources from \citet{UVplane2020}. The location of the quasar HE 0413-4031 is marked with a triangle.}
         \label{fig:UV_plane}
   \end{figure}
%-----------------------------------------------------------------

The location of the quasar HE 0413-4031 in the UV plane together with the sample from \citet{UVplane2020} is shown in Figure~\ref{fig:UV_plane} with a triangle colored with EW(MgII) as the rest of the sample. The UV FeII emission seems rather weak in the studied object. In the whole sample of \citet{UVplane2020}, there are only 14 quasars with $R_{Fe} <2.0$, and two with $R_{Fe} <1.5$. The number of sources with $R_{Fe} <2.37 $ is 47 (out of 2962 total). The smallest value in this sample is $1.28 \pm 0.15$, for the object SDSS 100352.65+212851.5 with the FWHM (MgII)$=5760\, {\rm km\, s^{-1}}$.

\section{Conclusions}
\label{sec:conclusions}

Our detailed study of the bright, intermediate-redshift quasar HE 0413-4031 based on 10-year monitoring with SALT combined with a single IR exposure with the telescope SOAR showed that
\begin{itemize}
\item determination of the redshift from a narrow emission line is important for the spectral decomposition of the entire IR/optical/UV band. The {\rm precise redshift determination of $z=1.39117 \pm 0.00017$} is significantly larger with respect to the previous value ($z=1.3764$),
\item the decomposition of the MgII+FeII complex based on the {\rm precise} redshift results only in the small blueshifted offset of the MgII line in the range between $-240$ and $-270\,{\rm km\,s^{-1}}$, which is consistent with the quasar composite spectra of the UV region,
\item  the velocity of the MgII line with respect to the systemic redshift is therefore significantly affected by the new redshift and the corresponding template since previously the velocity shift was in excess of $1000\,{\rm km\,s^{-1}}$,
\item some properties of the strong UV line, in particular MgII, such as its FWHM, EW, and the time delay with respect to the continuum are determined robustly, even without {\rm precise} redshift determination, so applications of MgII line for cosmology are justified,
\item the MgII emission based on the best-fitting KDP15 template with FWHM(MgII)$=4503.1\,{\rm km\,s^{-1}}$ is significantly correlated with the continuum. The final MgII time delay in the rest frame of the source, which is based on 7 different time-delay methods, is $\tau_{\rm MgII}=224^{+21}_{-23}$ days,
\item determination of the FeII properties, including its time delay with respect to the continuum, is very sensitive to the proper redshift determination and the use of the proper FeII template. The FeII emission is also strongly correlated with the continuum (the peak correlation coefficient is $r=0.86$), but the peak time delay is consistent with zero, which suggests that the FeII light curve is strongly contaminated with the continuum long-term trend,
\item after subtracting the long-term trend from both continuum and FeII light curves, we detect the candidate FeII time delay at $\tau=251^{+9}_{-7}$ days in the rest frame, albeit the correlation coefficient is relatively low ($r=0.27$),
\item the rest-frame time delay of UV FeII is consistent with the MgII/FeII FWHMs inferred using the FeII best-fit template KDP15. Assuming that the MgII/FeII-line emitting material is fully virialized, the FeII time delay should be $\tau_{\rm FeII}=\tau_{\rm MgII}(\text{FWHM}_{\rm MgII}/\text{FWHM}_{\rm FeII})^2\sim 257$ days, which is consistent with the reverberation result. Hence, for HE 0413-4031, the UV FeII material is more distant than the MgII line-emitting material by $\sim 27$ light days,
\item we update the MgII radius-luminosity relation with new measurements, which results in 194 sources, that is more than twice as much as used previously. The MgII $R-L$ relation is significantly flatter than the UV FeII, optical FeII, and H$\beta$ $R-L$ relations,
\item we suggest to build new FeII emission templates based on the newest version of the photoionization code \texttt{CLOUDY} since there are new FeII transitions included that indicate extension of the emission towards shorter wavelengths with respect to the older FeII templates currently in usage.
\end{itemize}

\begin{acknowledgements}
We thank the anonymous referee for constructive comments, which helped to improve the manuscript. All spectroscopic observations reported in this paper were obtained with the Southern African Large Telescope (SALT). The project is based on observations made with the SALT under programs 2012-2-POL-003, 2013-1-POL-RSA-002, 2013-2-POL-RSA-001, 2014-1-POL-RSA-001, 2014-2-SCI-004, 2015-1-SCI-006, 2015-2-SCI-017, 2016-1-SCI-011, 2016-2-SCI-024, 2017-1-SCI-009, 2017-2-SCI-033, 2018-1-MLT-004 (PI: B. Czerny). Polish participation in SALT is funded by grant No. MEiN nr 2021/WK/01. Based in part on observations obtained at the Southern Astrophysical Research (SOAR) telescope, which is a joint project of the Minist\'{e}rio da Ci\^{e}ncia, Tecnologia e Inova\c{c}\~{o}es (MCTI/LNA) do Brasil, the US National Science Foundation’s NOIRLab, the University of North Carolina at Chapel Hill (UNC), and Michigan State University (MSU). The project was partially supported by the Polish Funding Agency National Science Centre, project 2017/26/A/ST9/00756 (MAESTRO 9). BC and MZ acknowledge the OPUS-LAP/GAČR-LA bilateral project (2021/43/I/ST9/01352/OPUS 22 and GF23-04053L). This project has received funding from the European Research Council (ERC) under the European Union’s Horizon 2020 research and innovation program (grant agreement No. [951549]). SP acknowledges the financial support from the Conselho Nacional de Desenvolvimento Científico e Tecnológico (CNPq) Fellowships (164753/2020-6 and 313497/2022-2). MS acknowledges support from the Polish Funding Agency National Science Centre, project 2021/41/N/ST9/02280 (PRELUDIUM 20), the European Research Council (ERC) under the European Union’s Horizon 2020 research and innovation program (grant agreement 950533), and the Israel Science Foundation, (grant 1849/19). M.L.M.-A. acknowledges financial support from Millenium Nucleus NCN19-058 (TITANs). NW acknowledges the financial support from the GA\v{C}R EXPRO grant no GX21-13491X. We are grateful to Marianne Vestergaard for allowing us to use their FeII template.
\end{acknowledgements}

\begin{appendix}

\section{MgII time-delay determinations}
\label{sec_appendix_time_delay}

\subsection{Interpolated cross-correlation function}

\begin{figure}[h!]
    \centering
    \includegraphics[width=\columnwidth]{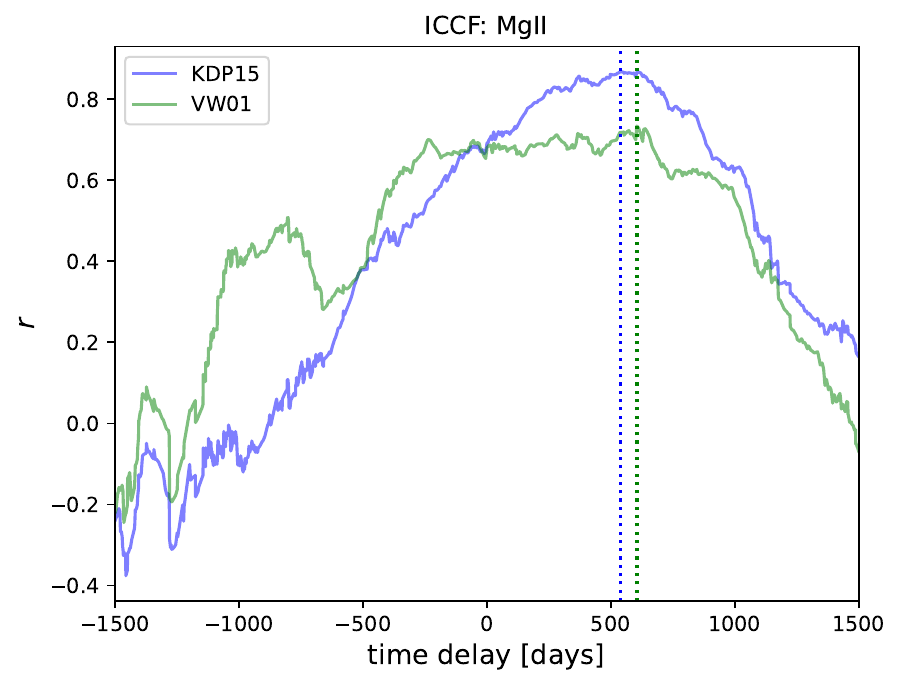}
    \caption{ICCF for the MgII emission using the KDP15 and the VW01 templates depicted by blue and green lines, respectively. The dotted vertical lines show the time-delay peaks at $541$ and $606$ days (highest cross-correlation coefficient). }
    \label{fig_iccf_MgII}
\end{figure}

We calculate the Interpolated cross-correlation function (ICCF) using the basic definition of the cross-correlation coefficient between two light-curve points separated by the time delay $\tau_{\rm k}=k\Delta t$,
\begin{equation}
    CCF(\tau_{k})=\frac{(1/N)\sum_{i=1}^{N-k}(x_i-\overline{x})(y_{i+k}-\overline{y})}{[(1/N)
    \sum_{i}^{N}(x_i-\overline{x})^2]^{1/2}[(1/N)\sum_i^{N}(y_i-\overline{y})^2]^{1/2}}\,,
    \label{eq_CCF}
\end{equation}
where $\Delta t=t_{i+1}-t_{i}$ is the regular time step, and $\overline{x}$ and $\overline{y}$ denote light curve means. Both continuum and line-emission light curves are interpolated with the time step of $\Delta t=1$ day (symmetric interpolation).
%For the MgII light curve, we first searched in the broad time-delay interval (-1500, 1500) days. We found that the peak can be located in the positive part at $\sim 500-600$ days, so we narrowed the interval to $(0, 1000)$ days. We used the ICCF as implemented in the \texttt{PyCCF} code of \citet{Sun2018}, based on the original analysis of \citet{Peterson1998}. Using the code, we also performed 1000 Monte Carlo simulations of the combined flux randomization and random subset selection. Based on the time-delay distributions, we inferred the peak, the centroid, and their corresponding uncertainties. 
In Figure~\ref{fig_iccf_MgII}, we show the ICCF for the MgII line emission inferred using the KDP15 template and the VW01 template. 

\subsection{Discrete correlation function (DCF)}

\begin{figure}
    \centering
    \includegraphics[width=\columnwidth]{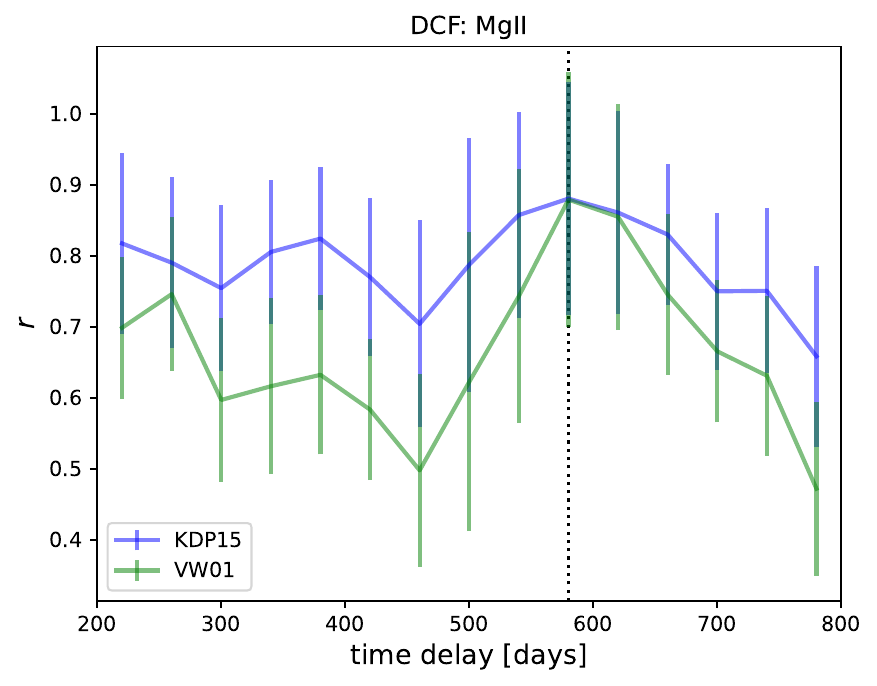}
    \caption{DCF for the MgII emission using the KDP15 and the VW01 templates depicted by blue and green lines, respectively. The dotted vertical line shows the time-delay peak at $580$ days (highest correlation coefficient). }
    \label{fig_dcf_MgII}
\end{figure}

We employ the discrete correlation function (DCF) that is suited for irregular and unevenly sampled light curves \citep{Edelson1988}. We utilize the \texttt{python} implementation of the DCF by \citet{Robertson2015}.
%We focus on the interval between $200$ and $800$ days and set the timestep to $40$ days, which is comparable to the mean sampling of the continuum light curve ($\sim  37.25$ days). Within the time bin, we apply the Gaussian weighting of light-curve pairs. 
For both KDP15 and VW01 FeII templates, we show the DCF results in Figure~\ref{fig_dcf_MgII}.

%The time-delay peak and its uncertainty are inferred from 500 bootstrap simulations, i.e. we generated 500 randomized light-curve subsets of both the continuum and the line emission. Subsequently, we fit a Gaussian function to the histogram of individual best time-delay peaks. The mean and the standard deviation are then determined from the best-fit values.

\subsection{$z$-transferred DCF (zDCF)}

\begin{figure}
    \centering
    \includegraphics[width=\columnwidth]{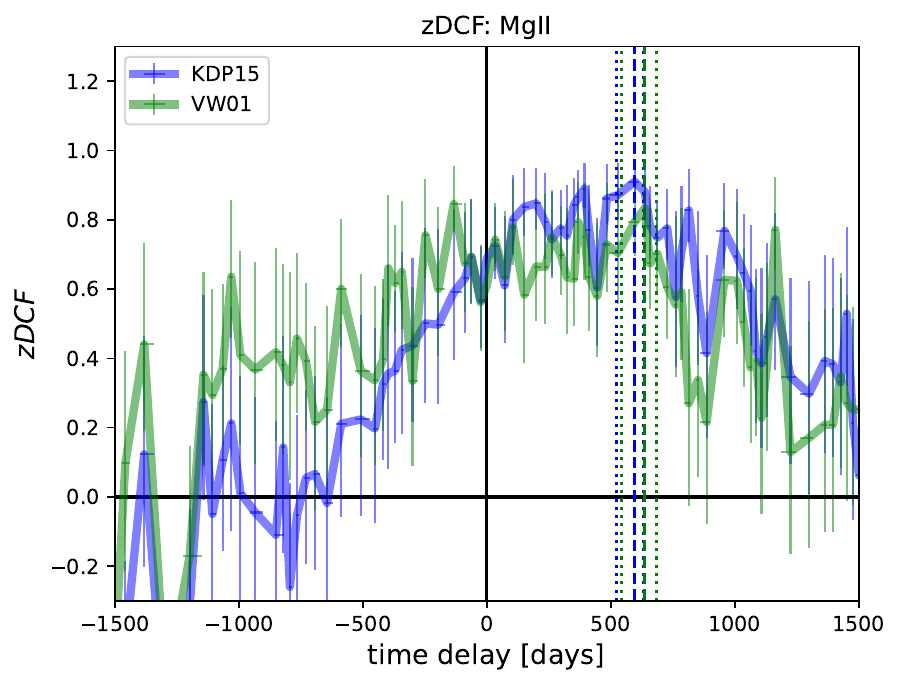}
    \caption{zDCF for the MgII emission using the KDP15 and the VW01 templates depicted by blue and green lines, respectively, with the corresponding error bars. The dashed vertical lines show the time-delay peaks at $597$ days (KDP15) and $638$ days (VW01) while the dotted lines mark the corresponding 1$\sigma$ uncertainties. }
    \label{fig_zdcf_MgII}
\end{figure}

The $z$-transferred DCF replaces equidistant time bins with equipopulation time bins. The minimum number of overlapping light-curve pairs is 12. The method was developed by \citet{1997ASSL..218..163A}, who also implemented it within a \texttt{fortran} code.
The zDCF is suitable for sparse and irregular light curves. %The uncertainties for each time delay are determined from 1000 Monte Carlo simulations using the flux-randomization technique.
In Figure~\ref{fig_zdcf_MgII}, we show zDCF as a function of the observed time delay for both KDP15 and VW01 FeII templates that are depicted by blue and green lines, respectively.
%Using the likelihood maximization, we determine the time-delay peaks and the corresponding 1$\sigma$ uncertainties. Within the uncertainties, both time-delay peaks at 597 and 638 days (in the observer's frame) are in agreement.

\subsection{$\chi^2$ method}

\begin{figure}
    \centering
    \includegraphics[width=\columnwidth]{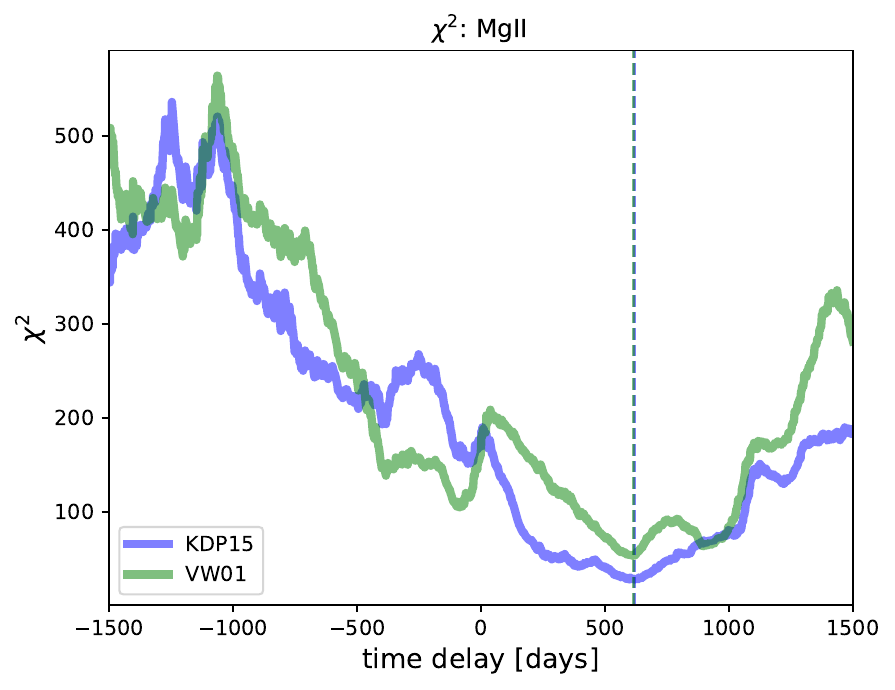}
    \caption{$\chi^2$ as a function of a time delay (in the observer's frame) for the MgII emission using the KDP15 and the VW01 templates depicted by blue and green lines, respectively. The dashed vertical lines depict the time-delay peaks at $619$ days (KDP15) and $616$ days (VW01). }
    \label{fig_chi2_MgII}
\end{figure}

We also apply the $\chi^2$ method frequently used in lensing studies. \citet{Czerny2013} showed that this method performs better than the ICCF for red-noise AGN variability.
%A linear interpolation is employed to first interpolate the continuum with respect to line emission (asymmetric interpolation), and then vice versa. The final $\chi^2$ is then calculated as the mean of both interpolations and it serves as a proxy of the light-curve similarity. Hence, the minimum $\chi^2$ value corresponds to the time delay between the continuum and line-emission light curves.
In Figure~\ref{fig_chi2_MgII}, we show $\chi^2$ as a function of the time delay of the MgII line emission with respect to the photoionizing continuum. The solid blue line stands for the KDP15 FeII template, while the solid green line represents the VW01 template. 
%The vertical dashed blue and green lines depict the $\chi^2$ minima. The difference is hardly noticeable since the minimum $\chi^2$ for the KDP15 template is at 619 days in the observer's frame, while it is at 616 days for the VW01 template.

\subsection{JAVELIN}

\begin{figure*}
    \centering
    \includegraphics[width=\columnwidth]{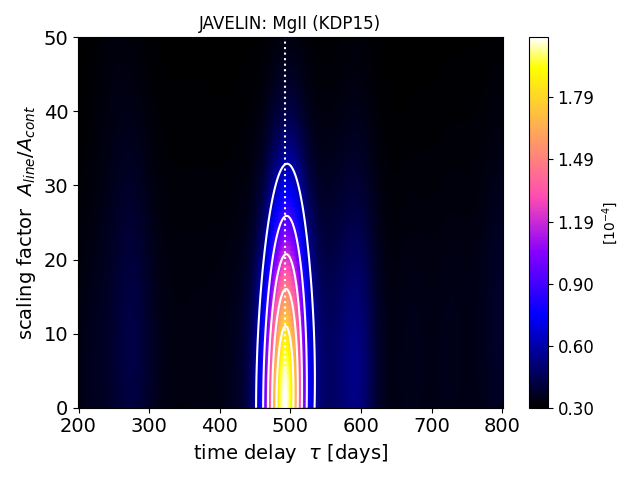}
    \includegraphics[width=\columnwidth]{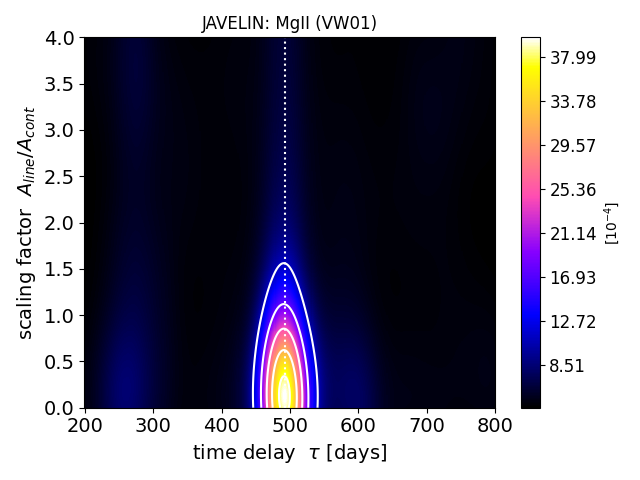} 
    \caption{Heat map with contours for the posterior distributions of the scaling factor vs. the time delay (in the observer's frame) as determined by the JAVELIN method for the MgII emission light curve. We performed the analysis using the KDP15 FeII template (left panel) and the VW01 FeII template (right panel). The dashed vertical line depicts the time-delay peak at $492.5$ days, which is the same for both FeII templates. The color scale depicts the Gaussian kernel density of MCMC realizations (altogether 40\,000) in the scaling factor--time delay plane with the grid of 400 $\times$ 400 bins. }
    \label{fig_javelin_MgII}
\end{figure*}

Another method to estimate the time delay in galactic nuclei is to model the continuum emission as a stochastic process, for example by using a damped random walk \citep{Kelly2009,2010ApJ...708..927K,2010ApJ...721.1014M, kozlowski2016}, and to assume that the line emission is a time-delayed, scaled, and smoothed version of the continuum emission. Based on this approach the JAVELIN package (Just Another Vehicle for Estimating Lags In Nuclei) was developed and tested \citep{Zu2011,Zu2013,2016ApJ...819..122Z}.  
%JAVELIN first infers the posterior probability distribution of the continuum variability timescale and amplitude. Then by applying the Markov Chain Monte Carlo, it infers the posterior probability distributions of the time lag, the smoothing width of the top-hat function, and the scaling factor (the ratio of the line emission and the continuum amplitudes, $A_{\rm line}/A_{\rm cont}$). 
We applied JAVELIN to infer independently the time delay of the MgII emission with respect to the continuum. In Figure~\ref{fig_javelin_MgII}, we plot the heatmaps of the scaling factor vs. time delay. In the left panel, we present the case with the KDP15 FeII template, while in the right panel, we show the case with the VW01 FeII template.
%In both cases, the JAVELIN MCMC analysis was performed in the time range between 200 and 800 days (peaks outside this range are rather small and spurious) with in total of 40\,000 burn-in iterations. In both cases, the peak of the posterior time-delay distribution is well-defined, it is at $493^{+4}_{-15}$ days for the KDP15 template and at $493^{+3}_{-17}$ days for the VW01 template. 

\subsection{Measures of regularity and randomness}

Recently, the estimators of time-series regularity or randomness were proposed as a way to estimate the true time delay between the two light curves \citep{2017ApJ...844..146C}. %In comparison with the ICCF, they do not require polynomial interpolation, nor is it necessary to perform binning in the correlation space as for DCF and zDCF methods. In addition, no AGN variability process, such as a red noise or a damped random walk, is assumed to fit the AGN continuum, such as for the JAVELIN algorithm. 
%\citet{2017ApJ...844..146C} showed that the best measure of time-series regularity is an optimized von Neumann scheme that works with the combined light curve in the form $F(t,\tau)=\{(t_i,f_i)\}_{i=1}^{N}=F_1 \cup F_2^{\tau}$, where $F_1$ is the continuum light curve and $F_2^{\tau}$ is a time-shifted line-emission light curve. Then the von Neumann estimator can be defined as the mean square successive difference of the combined light curve,
%\begin{equation}
%    E(\tau)=\frac{1}{N-1} \sum_{i=1}^{N-1} [F(t_i)-F(t_{i+1})]^2\,,
%    \label{eq_vn_estimator}
%\end{equation}
%whose global minimum $E_{\rm min}(\tau')$ indicates the time delay $\tau'$ that can be considered as the candidate actual time delay between the continuum and the line-emission light curves, i.e. for such a time delay the combined light curve is the most regular.
\begin{figure}
    \centering
    \includegraphics[width=\columnwidth]{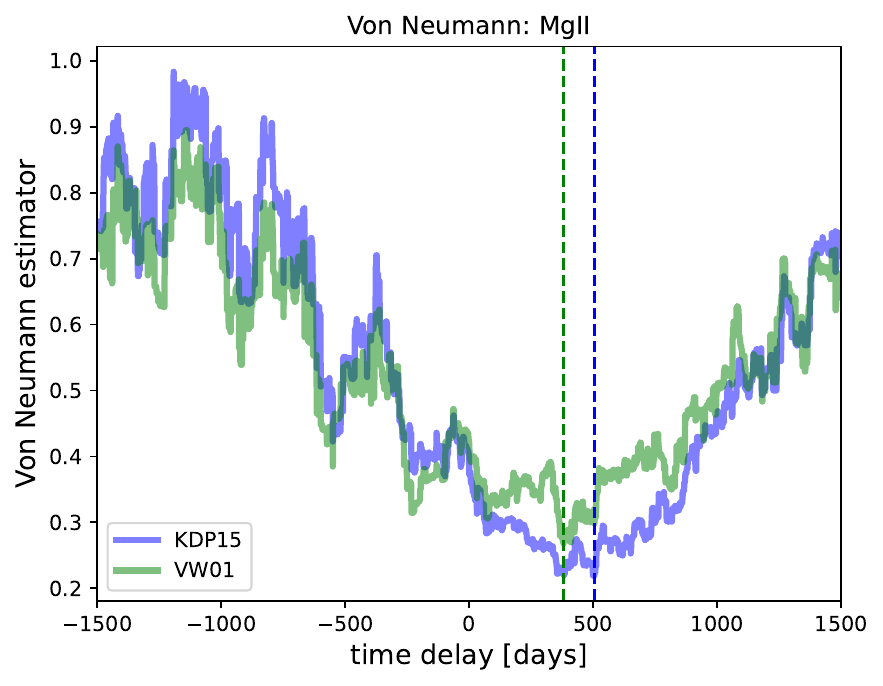}
    \caption{Von Neumann estimator as a function of a time delay (in the observer's frame) for the MgII emission using the KDP15 and the VW01 templates depicted by blue and green lines, respectively. The dashed vertical lines depict the time delays corresponding to the estimator minima at $504$ days (KDP15) and $380$ days (VW01), respectively. }
    \label{fig_vn_MgII}
\end{figure}
In Figure~\ref{fig_vn_MgII}, we present the von Neumann estimator as a function of the time delay in the observer's frame of reference. As was done previously for the other methods, two FeII templates are compared: KDP15 (blue line) and VW01 (green line). 
%There is a shift of about 120 days in terms of the estimator minimum. For the KDP15 template, the minimum is at 504 days while for the VW01 template, it is at 380 days. The peak time delays and their uncertainties for both templates are inferred from the time-delay values corresponding to the von Neumann estimator minima for 1000 light-curve pairs generated using a bootstrap method.
Furthermore, we also calculate the Bartels estimator, which is similar to the von Neumann estimator but rather operates on the ranked combined light curve. %, $F_{\rm R}(t,\tau)$. 
 In Figure~\ref{fig_bartels_MgII}, we show the Bartels estimator values as a function of the time delay for both templates. Overall, the minima values and their uncertainties as inferred from the best time-delay distributions are quite similar to the values determined using the von Neumann estimator.

\begin{figure}
    \centering
    \includegraphics[width=\columnwidth]{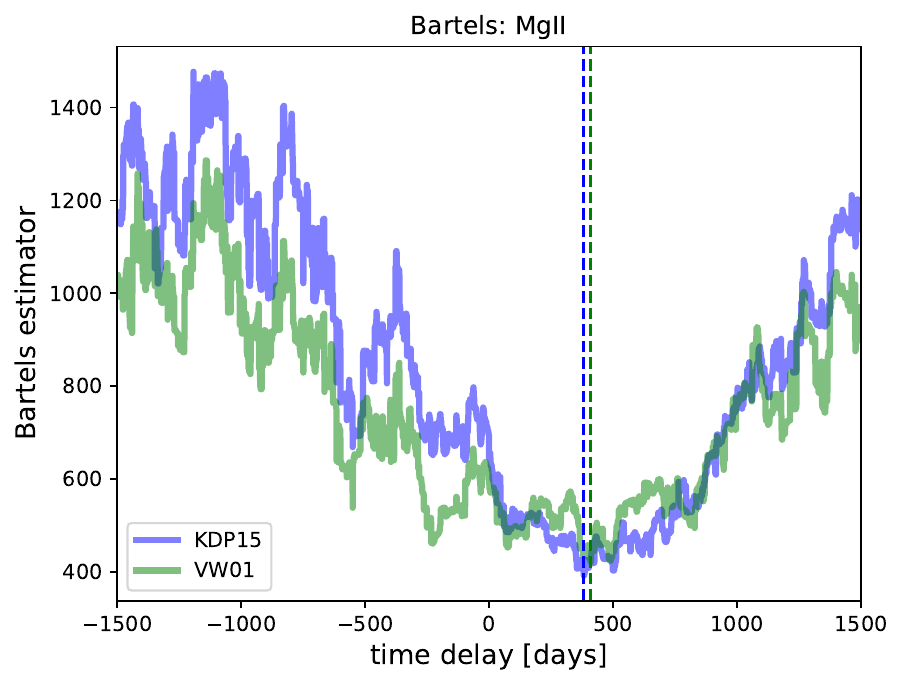}
    \caption{Bartels estimator as a function of a time delay (in the observer's frame) for the MgII emission using the KDP15 and the VW01 templates depicted by blue and the green lines, respectively. The dashed vertical lines depict the time delays corresponding to the estimator minima at $382$ days (KDP15) and $410$ days (VW01), respectively. }
    \label{fig_bartels_MgII}
\end{figure}

\subsection{Subtraction of long-term trend}
\label{subsec_sub}

In a similar way as for the UV FeII line (see Subsection~\ref{subsec_time_delay}), we verify that for the MgII light curve the long-term (LT) trend subtraction does not obliterate the main time-delay peak.
\begin{figure}
    \centering
    \includegraphics[width=\columnwidth]{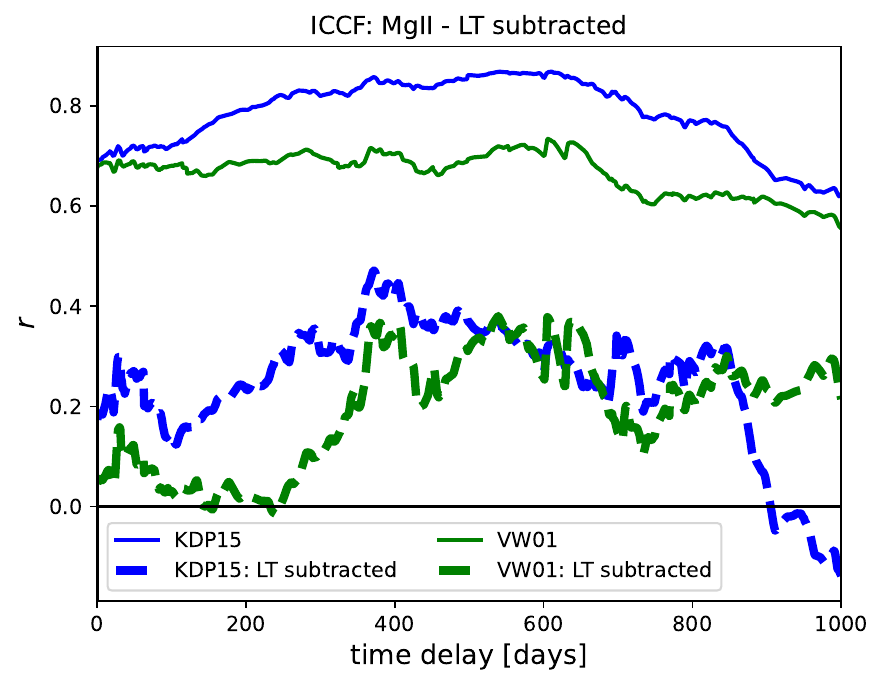}
    \caption{ICCF as a function of the time delay in the observer's frame for the MgII light curve with the LT trend subtraction. Solid lines correspond to the original light curves (blue for the KDP15 template and green for the VW01 template). Dashed lines correspond to the continuum and the MgII light curves with the LT trend subtracted using the third-order polynomial to represent the LT trend (a blue dashed line stands for the KDP15 template, while the green dashed line corresponds to the VW01 template). Although a significant drop in the correlation coefficient is noticeable, the main time-delay peak around $\sim 400-600$ days can be recovered.}
    \label{fig_MgII_LT_effect}
\end{figure}
We fit the third-order polynomial to both continuum and MgII light curves. Subsequently, we subtract the best-fit trend from both light curves. Then we apply the ICCF, for which we find the time delays corresponding to the peak $r$ value, the centroid, and the peak distributions. Both KDP15 and VW01 templates are considered. The results are summarized in Table ~\ref{tab_mgII_sub}.

\begin{table}[h!]
    \centering
     \caption{Summary of MgII emission time delays inferred using the ICCF for the light curves with the LT trend subtracted. Time delays are expressed in days in the observer's frame, while the last two rows represent the mean time delays expressed in days in the rest frame of the source.}
    \begin{tabular}{c|c|c}
    \hline
    \hline
    Method   & KDP15 & VW01  \\
    \hline
    ICCF* - highest $r$ & 372 ($r=0.47$) &  539 ($r=0.38$)  \\
    ICCF* - peak     & $ 410^{+345}_{-50}$   & $605^{+242}_{-238}$  \\
    ICCF* - centroid & $ 446^{+317}_{-103}$    &  $571^{+275}_{-203}$   \\
    \hline
    $\overline{\tau}$ -- observer's frame & $428^{+234}_{-57}$   &  $588^{+183}_{-156}$    \\
     $\overline{\tau}$ -- rest frame    & $179^{+98}_{-24}$   &  $246^{+77}_{-65}$    \\
       \hline
    Final MgII time delay & \multicolumn{2}{c}{$213^{+62}_{-35}$} \\
    \hline
    \end{tabular}   
    \label{tab_mgII_sub}
\end{table}

It is possible to recover the time delay at $\sim 400-600$ days in the observer's frame, although with a significantly lower correlation of $r\sim 0.38-0.47$. For the KDP15 template, the peak and centroid time delays are smaller than the previously determined values, though within the uncertainties they are consistent with the time delays reported without any subtraction. When we average the peak and the centroid values for both templates, the rest-frame time delay is $213^{+62}_{-35}$ days, which is in agreement with the previously reported value.

\end{appendix}
\end{document}